%
%
%
%
%
%
%
%
%
%
%
%
%
%
\input phyzzx.tex
\input epsf
%
%
\catcode`\@=11 
\def\papersize{\hsize=40pc \vsize=53pc \hoffset=0pc \voffset=1pc
   \advance\hoffset by\HOFFSET \advance\voffset by\VOFFSET
   \pagebottomfiller=0pc
   \skip\footins=\bigskipamount \normalspace }
\catcode`\@=12 
\papers
\vsize=23.cm
\hsize=15.cm
\newcount\figno
\figno=0
\def\fig#1#2#3{
\par\begingroup\parindent=0pt\leftskip=1cm\rightskip=1cm\parindent=0pt
\baselineskip=11pt
\global\advance\figno by 1
\midinsert
\epsfxsize=#3
\centerline{\epsfbox{#2}}
\vskip 12pt
{\bf Fig. \the\figno:} #1\par
\endinsert\endgroup\par
}
\def\figlabel#1{\xdef#1{\the\figno}}
\def\encadremath#1{\vbox{\hrule\hbox{\vrule\kern8pt\vbox{\kern8pt
\hbox{$\displaystyle #1$}\kern8pt}
\kern8pt\vrule}\hrule}}
%
\vsize=23.cm
\hsize=15.cm
\def\IM{\mathop{\Im m}\nolimits}

\def\Z{{\bf Z}}
\def\ZZ{$\Z_2$}
\def\R{{\bf R}}

\def\C{${\cal C}$}
\def\CP{${\cal C}^+$}
\def\CM{${\cal C}^-$}
\def\CO{${\cal C}^0$}

\def\pp{$(n_e,n_m)$}
\def\to{\rightarrow}

\def\ad{a_D}
\def\adu{a_D^{(1)}}
\def\add{a_D^{(2)}}
\def\adt{a_D^{(3)}}
\def\au{a^{(1)}}
\def\adeux{a^{(2)}}
\def\at{a^{(3)}}
\def\e{\epsilon}

\def\wt{w^{(3)}}
\def\wu{w^{(1)}}

\tolerance=500000
\overfullrule=0pt

\Pubnum={LPTENS-96/22 \cr
HUB-EP-96/11\cr
{\tt hep-th@xxx/9605101} \cr
May 1996}

\date={}
\pubtype={}
\titlepage
\title{{\bf Curves of Marginal Stability, and Weak and\break
Strong-Coupling BPS Spectra in $N=2$ Supersymmetric QCD}
}
\author{Adel~Bilal
\foot{Work partially carried out while the author was visiting the
Institut f\"ur Physik, Humboldt Universit\"at Berlin,
Invalidenstra\ss e 110, D-10115 Berlin}
}
\andauthor{Frank~Ferrari}
\vskip .5cm
\address{
CNRS - Laboratoire de Physique Th\'eorique de l'\'Ecole
Normale Sup\'erieure
\foot{{\rm unit\'e propre du CNRS, associ\'ee \`a l'\'Ecole Normale
Sup\'erieure et l'Universit\'e Paris-Sud}}    \break
24 rue Lhomond, 75231 Paris Cedex 05, France  \break
{\tt bilal@physique.ens.fr,\   ferrari@physique.ens.fr}
}

\vskip 0.5cm
\abstract{We explicitly determine the global structure of the
$SL(2,\Z)$ bundle over the Coulomb branch of the moduli space 
of asymptotically free $N=2$
supersymmetric Yang-Mills theories with gauge group
$SU(2)$ when massless hypermultiplets
are present. For each relevant number of flavours, we show that
there is a curve of marginal stability on the Coulomb branch,
diffeomorphic to a circle, across which the BPS spectrum is
discontinuous. We determine rigorously and completely the BPS
spectra inside and outside the curve. In all cases, the spectrum
inside the curve consists of only those BPS states that are 
responsible for the singularities of the low energy effective action
(in addition to the massless abelian gauge multiplet which is
always present). The predicted decay patterns across the curve
of marginal stability are perfectly consistent with all quantum
numbers carried by the BPS states. As a byproduct, we also show
that the electric and magnetic
quantum numbers of the massless states
at the singularities proposed by Seiberg 
and Witten are the only possible ones.

%

\endpage
\pagenumber=1

 \def\PL #1 #2 #3 {Phys.~Lett.~{\bf #1} (#2) #3}
 \def\NP #1 #2 #3 {Nucl.~Phys.~{\bf #1} (#2) #3}
 \def\PR #1 #2 #3 {Phys.~Rev.~{\bf #1} (#2) #3}
 \def\PRL #1 #2 #3 {Phys.~Rev.~Lett.~{\bf #1} (#2) #3}
 \def\CMP #1 #2 #3 {Comm.~Math.~Phys.~{\bf #1} (#2) #3}
 \def\IJMP #1 #2 #3 {Int.~J.~Mod.~Phys.~{\bf #1} (#2) #3}
 \def\JETP #1 #2 #3 {Sov.~Phys.~JETP.~{\bf #1} (#2) #3}
 \def\PRS #1 #2 #3 {Proc.~Roy.~Soc.~{\bf #1} (#2) #3}
 \def\JFA #1 #2 #3 {J.~Funkt.~Anal.~{\bf #1} (#2) #3}
 \def\LMP #1 #2 #3 {Lett.~Math.~Phys.~{\bf #1} (#2) #3}
 \def\IJMP #1 #2 #3 {Int.~J.~Mod.~Phys.~{\bf #1} (#2) #3}
 \def\FAA #1 #2 #3 {Funct.~Anal.~Appl.~{\bf #1} (#2) #3}
 \def\AP #1 #2 #3 {Ann.~Phys.~{\bf #1} (#2) #3}
 \def\MPL #1 #2 #3 {Mod.~Phys.~Lett.~{\bf #1} (#2) #3}

\def\rw{$\, {\cal R}_W$}
\def\rs{$\, {\cal R}_S$}
\def\rsp{$\, {\cal R}_{S+}$}
\def\rsm{$\, {\cal R}_{S-}$}
\def\rso{$\, {\cal R}_{S0}$}
\def\sw{$\, {\cal S}_W$}
\def\ss{$\, {\cal S}_S$}
\def\ssp{$\, {\cal S}_{S+}$}
\def\ssm{$\, {\cal S}_{S-}$}
\def\sso{$\, {\cal S}_{S0}$}

\REF\FB{F. Ferrari and A. Bilal, {\it The strong-coupling spectrum of
Seiberg-Witten theory}, Nucl. Phys. {\bf B469} (1996) 387,
{\tt hep-th/9602082}.}

\REF\KL{A. Klemm, W. Lerche, P. Mayr, C. Vafa and N. Warner,
{\it Self-dual strings and $N=2$ supersymmetric field theory},
preprint CERN-TH/96-95, HUTP-96/A014, USC-96/008,
{\tt hep-th/9604034}.}

\REF\LIN{U. Lindstr\"om, M. Ro\v cek, {\it A note on the
Seiberg-Witten solution of $N=2$ super Yang-Mills theory},
\PL B355 1995 492 , {\tt hep-th/9503012}.}

\REF\FAY{A. Fayyazuddin, {\it Some comments on $N=2$ supersymmetric
Yang-Mills}, \MPL A10 1995 2703 , {\tt hep-th/9504120}.}

\REF\SW{N. Seiberg and E. Witten, {\it Electric-magnetic duality, monopole
condensation, and confinement in $N=2$ supersymmetric Yang-Mills theory}, 
\NP B426 1994 19 , {\tt hep-th/9407087}.}

\REF\SWII{N. Seiberg and E. Witten, {\it Monopoles, duality and chiral
symmetry breaking in $N=2$ supersymmetric QCD}, \NP B431 1994 484 ,
{\tt hep-th/9408099}.}

\REF\VAF{S. Cecotti, P. Fendley, K. Intriligator and C. Vafa,
{\it A new supersymmetric index}, \NP B386 1992 405 ;\hfill\break
S. Cecotti and C. Vafa, {\it On classification of $N=2$
supersymmetric theories}, \CMP 158 1993 569 .}

\REF\HEN{M. Henningson, {\it Discontinuous BPS spectra in $N=2$ gauge theory}, \NP B461
1996 101 , {\tt hep-th/9510138}.}

\REF\SEN{A. Sen, {\it Dyon-monopole bound states, self-dual harmonic
forms on the multi-monopole moduli space, and $SL(2,\Z)$ invariance
in string theory}, \PL B329 1994 217 , {\tt hep-th/9402032}.}

\REF\SET{S. Sethi, M. Stern and E. Zaslow, {\it Monopole and
dyon bound states in $N=2$ supersymmetric Yang-Mills theories},
\NP B457 1995 484 , {\tt hep-th/9508117}.}

\REF\IY{K. Ito and S.-K. Yang, {\it Prepotentials in N=2 SU(2)
supersymmetric Yang-Mills theory with massless hypermultiplets}, \PL
B366 1996 165 , {\tt hep-th/9507144}.}

\REF\SWREV{A. Bilal, {\it Duality in $N=2$ susy $SU(2)$ Yang-Mills theory: 
A pedagogical introduction to the work of Seiberg and Witten},  
\'Ecole Normale Sup\'erieure preprint LPTENS-95/53, {\tt hep-th/9601007}.}

\REF\ERD{A. Erdelyi et al, {\it Higher Transcendental Functions}, Vol 1, McGraw-Hill,
New York, 1953.}

\REF\ARG{P.C. Argyres, A.E. Faraggi and A.D. Shapere, {\it Curves
of marginal stability in $N=2$ super-QCD}, preprint IASSNS-HEP-94/103, 
UK-HEP/95-07, {\tt hep-th/9505190};\hfill\break
M. Matone, {\it Koebe $1/4$-theorem and inequalities in
$N=2$ super-QCD}, \PR D53 1996 7354 , {\tt hep-th/9506181}.}

\REF\GAU{J.P. Gauntlett and J.A. Harvey, {\it S-duality and the
dyon spectrum in $N=2$ super Yang-Mills theory},
\NP B463 1996 287 , {\tt hep-th/9508156}.}

\chapter{Introduction and Summary}

Recently, some new methods have been imagined which led to the complete
determination of the  spectrum of BPS states\foot{
By BPS spectrum, we always mean the set of BPS quantum states
existing in the theory, not the mass spectrum.}
in $N=2$ supersymmetric theories (first in [\FB] and later in [\KL ]). 
Part of these results had been anticipated at the conjectural level
[\LIN ,\FAY ].
The problem was open since the
work of Seiberg and Witten [\SW ,\SWII] where it was realized that
there may exist some regions in the moduli space where the BPS
spectrum cannot be continuously related to the semi-classical one.
This stems from the fact that the duality transformations 
(monodromies) involved in the description of the low energy physics
of these theories are not quantum symmetries at all  energy scales.
In particular, the semi-classical BPS spectrum need not to be invariant under the
full monodromy group, but only under a subgroup of it,
generated by the monodromy at infinity
$M_\infty$, which shifts
the electric charge of a dyon by an integer multiple of its magnetic charge.
The discontinuity in the BPS spectrum is possible because
there exists a curve of marginal stability where usually stable
BPS particles become degenerate in mass with other BPS states.
This phenomena was first highlighted in two dimensional theories
in [\VAF ].
As for the pure $N_f =0$ gauge theory,  we will show here
that in the cases $N_f=1,2,3$ 
these curves 
separate what we call a strong-coupling from a weak-coupling
region. The latter name is used for the region that contains the semi-classical
one, where the Higgs vacuum expectation value goes to infinity and the theories
are weakly coupled. On the other hand, this so-called weak-coupling region
extends all the way to the curve of marginal stability where the physics is
actually strongly coupled. The existence of this curve is a genuine
non-perturbative effect in the theories with zero bare masses.\foot{
Note, however, that for large bare masses of the quarks, such curves also exist at weak
coupling where jumps in the spectrum can be studied using semi-classical methods
[\HEN].}

In [\FB] the authors have shown that in the $N_f =0$ theory
the strong-coupling BPS spectrum 
must be invariant under some duality transformations
which {\it do not} belong to the monodromy group, but which are
related to the existence of a global exact quantum symmetry
acting on the moduli space. There it was a \ZZ\ symmetry
coming from the spontaneous breaking of a \Z $_8$ global symmetry
down to $\Z _4 $. Surprisingly, combining this symmetry
with the fact that the low energy physics cannot be described 
using a unique set of elementary light fields over the
whole moduli space (that is, the $SL(2,\Z)$ bundle is not trivial),
it was shown in [\FB] that, in the strong-coupling region, the BPS states must come in
{\it multiplets of the  broken symmetry}
(the ``\ZZ\ pairs''). This was at the basis of the proof that the
strong-coupling spectrum consists only of the particles
responsible for the singularities, namely the magnetic monopole 
$(n_e,n_m)=\pm (0,1)$ and the dyon of 
unit electric charge (in the normalisations of
[\FB,\SW]), alternatively
described as $\pm (\pm 1,1)$. 
Also included in [\FB] was a new way of determining the weak-coupling
spectrum, independent of the semi-classical approach. It was shown
that no state with a magnetic charge $|n_m| \geq 2$ can be present.
As noted by Sen [\SEN] this can also be viewed as a consequence
of the conjectured self-duality of the $N=4$ theory. The same
result also follows solely from semi-classical reasoning as noted
in [\SET].
The results of [\FB] have been subsequently confirmed in [\KL ]
where a completely independent method from string theory
was used.

In the present paper, we will extend and generalize the methods and
results of [\FB] to the case where $N_f=1,\, 2$ or $3$ 
hypermultiplets (quarks), without bare masses,
in the spin $1/2$ representation of the gauge group $SU(2)$,
are included.
If not for the bare masses set to zero, these are the most general
asymptotically free theories of this type.
For $N_f=1$, the moduli space of vacua ${\cal M}_1$
is homeomorphic to the compactified complex $u$-plane,
where $u=\langle $tr$\,\phi ^2 \rangle$ and $\phi $ is the
Higgs field in the adjoint representation of $SU(2)$.
This is a Coulomb branch on which the gauge group is broken down
to $U(1)$. For $N_f\geq 2$, in addition to a similar Coulomb branch,
the moduli space has also  Higgs branches on which the gauge group
is completely broken. In this paper we focus entirely on the Coulomb
branch where magnetically charged particles exist. 
We will denote this Coulomb branch as ${\cal M}_{N_f}$ in the 
following.
The 
expectation value $a(u)$ of the scalar field $\phi$ and  its dual
$\ad(u)$ 
are given by the two
periods of certain one-forms on appropriate elliptic curves [\SWII].
As such they satisfy Picard-Fuchs differential equations. These
Picard-Fuchs equations were derived in [\IY] where solutions, valid
in the neighbourhoods of the various singularities,
were also given. 
However, though this was enough to determine power series expansions
for the prepotential and its dual, no explicit expressions
valid on the whole Coulomb branch were given.
In the following, we will express $\ad $ and $a$ in terms of
standard hypergeometric functions and explicitly
display the global analytic structure
of the $SL(2,$\Z $)$ bundle $E_{N_f}$ over ${\cal M}_{N_f}$.
This will prove useful when discussing the strong-coupling
spectrum.

Once we have these solutions, it is indeed straightforward to
determine the curves of marginal stability where $\ad /a$ is
real. For each $N_f=1,\, 2,\, 3$, one has a single closed curve
diffeomorphic to a circle. Of course, since a BPS state can become
massless only at points on this curve, as discussed in
[\FB], the curve goes through all
singular points of the Coulomb branch (except $u=\infty$). For $N_f=2$, the
solutions $\ad(u),\ a(u)$, up to a multiplicative constant, 
and hence the curve are identical to those
of $N_f=0$. For $N_f=3$, the curve is a rescaled and shifted version
of the $N_f=0$ curve. 

To determine the spectra, we will in particular exploit the global
\Z $_{4-N_f}$ symmetry acting on the Coulomb branch of the moduli space. 
{}From this point
of view, the richest structure is for $N_f=1$ which exhibits a
\Z $_3$ symmetry. For $N_f =3$ there is no such symmetry, but
we will show that there it is not needed for the 
determination of the spectrum.
In this case, the strong-coupling spectrum consists of only the
dyon\foot{
Note that we changed the sign of the electric charge $n_e$ with
respect to the conventions of Seiberg and Witten [\SW,\SWII].}
$(-1,2)$ with magnetic charge two 
and the magnetic monopole $(0,1)$. 
The weak-coupling
spectrum contains the dyons $(n,1)$ and $(2n+1,2)$ for all integers
$n$, in
addition to the elementary quarks $(1,0)$ and
W bosons $(2,0)$, while we prove that no magnetic charges $\vert n_m\vert \ge 3$ 
can exist.
For $N_f=2$, we have only $(0,1)$ and $(1,1)$ in the strong-coupling
region, the semi-classical spectrum consisting of all the dyons
$(n,1)$ where $n$ is any integer, together with the quarks and
W bosons. We show that no $\vert n_m\vert \ge 2$ are allowed.
For $N_f=1$, 
the strong-coupling spectrum consists again of only the 
three states responsible for the singularities, namely the monopole
$(0,1)$ and the dyons $(-1,1)$ and $(-2,1)$.
For the weak-coupling spectrum we prove that again no
magnetic charges $\vert n_m\vert \ge 2$ can exist, while it contains all dyons
$(n,1),\ n\in \Z$, as well as the elementary quarks and W bosons.
Finally note that the neutral abelian $N=2$ vector multiplet 
is present in all cases over the whole moduli space.

We begin in Section 2 with a brief
overview of the results obtained in [\FB] for $N_f=0$ and 
then discuss some important facts and ideas we will use in this
work. There, we also  discuss in some detail the r\^ole of the broken global symmetries
and the relevant representations of the flavour groups.
Then, in Sections 3 to 5,
we present our results for respectively $N_f =1$, $2$ and $3$.
These sections are organized as follows: we start by presenting the
structure of the singularities
on the Coulomb branch, and explain why
it is unique.
The global analytic structure of the $SL(2,$\Z $)$ bundle $E_{N_f}$ 
is then determined, and 
the
curve ${\cal C}_{N_f}$ of marginal stability described. When it exists,
we also study the global quantum symmetry $\Z _{4-N_f}$
acting on ${\cal M}_{N_f}$.
This leads to the determination of the BPS spectra. We then 
show that the predicted decay reactions across the curve
${\cal C}_{N_f}$ are perfectly consistent with all quantum
numbers carried by the BPS states, and in particular with the branching rules of the
representations of the flavour groups.

\chapter{Review of $N_f=0$, and general considerations}

In this Section, we review or introduce certain results and ideas,
which will prove very useful in the
following.
Along the Coulomb branch,
the effective action is given in
terms of a prepotential ${\cal F}(a)$, and the knowledge of $a(u)$ and
$\ad(u)\equiv\displaystyle {1\over 2}{\partial {\cal F}\over
\partial a}(a(u))$ on all of the
Coulomb branch ${\cal M}$, which is the (compactified) $u$-plane, allows
one to reconstruct ${\cal F}(a)$. Also the mass of any BPS state with
electric and magnetic charges $n_e$ and $n_m$ is determined as\foot{
We will equally note the sections of $E_{N_f}$ by line or
column vectors.}
$$m=\sqrt{2}\,\big| a n_e - a_D n_m  \big| = 
\sqrt{2}\,\big| \eta \bigl( (n_e,n_m),(a_D ,a) \bigr)\big|\ ,
\eqn\di$$
where $\eta$ is the standard symplectic product.
In Seiberg and Witten's work [\SW,\SWII], the physical requirement of
having certain monodromies around singularities in the moduli space
is translated by equating $da_D / du$ and
$da / du$ with period integrals of the only holomorphic
one-form on an appropriate genus one Riemann surface. 
The monodromies are determined by the asymptotic behaviour of
$(a_D ,a)$ near the relevant singularities. Near the point at infinity 
the one loop $\beta $ function of the microscopic theory leads to
$$\eqalign{
a^{(N_f)}(u)&\sim {1\over 2}\ \sqrt{2u}\cr
\ad^{(N_f)}(u)&\sim {i\over 4 \pi}\, (4-N_f)\, \sqrt{2u}\, \log u\cr
}
\eqn\dii$$
and 
$$M_{\infty}=\pmatrix{-1 &\hfill 4-N_f \cr\hfill 0 &-1\cr}.
\eqn\diii $$
The singularities at finite $u$ are due to BPS states 
$(n_e ,n_m)$ becoming massless.
Using the $\beta $ function of the abelian low energy theory in which
$(n_e ,n_m)$ is coupled locally, 
one can obtain the corresponding monodromy matrix. When $d$
BPS hypermultiplets $(n_e ,n_m)$ become massless, which occurs
when the states carry a $d$ dimensional
representation of the flavour group, we have 
$$M_{(n_e ,n_m),d}=\pmatrix{1-n_e n_m\, d & n_e^2\, d \cr
-n_m^2\, d &1+ n_e n_m\, d\cr }.
\eqn\div $$

The easiest way of determining the explicit solution for
$(a_D ,a)$ is to use
the Picard-Fuchs equation satisfied by the period integrals.
These have been derived in
[\IY]
for the elliptic curves given in [\SWII], and are
second order differential equations having two linearly independent
solutions. The knowledge of the
asymptotics fixes 
the appropriate linear combinations for $a_D$ and $a$. This will
be worked out for $N_f=1$ and $N_f=3$ below.

We also recall that it follows from \di\ that for given $u$ the mass of 
any BPS state is proportional to the euclidean length of the vector
$n_e a - n_m \ad$ that lies on the lattice spanned by the two complex
numbers $a(u)$ and $\ad(u)$. Charge conservation and the triangle
inequality then imply that a state \pp\ can only decay if $n_e$ and
$n_m$ are not relatively prime, i.e. if $(n_e,n_m)=q(n,m)$ with
$n,m,q\in \Z,\ q\ne \pm 1$. Hence, BPS states with $n_e,n_m$
relatively prime are stable. This argument fails if 
$\ad / a \in\R$,
so that the lattice collapses onto a single
line. On the moduli space ${\cal M}_{N_f}$, the curve of marginal stability
\C $_{N_f}$ is defined as ${\cal C}_{N_f}=\Bigl\{ u\in {\cal M}\bigm|
\ad ^{(N_f)}/
a^{(N_f)}\in\R\Bigr\}$.
Hence, at any two points $u$ and $u'$ that can be
joined by a path that does not cross the curve \C\ the spectrum of stable BPS states
is necessarily the same.
As already mentioned in the introduction we call the BPS spectra outside and inside the
curve the weak and strong-coupling spectra, denoted \sw\ and \ss, and refer to the
regions as \rw\ and \rs. 

\section{Review of $N_f =0$}

For a pedagogical introduction to this case, see e.g. [\SWREV ].
We will use the normalization convention of [\SWII ] for 
the electric charge, so that $n_e$ is always an integer. 
Indeed, the Dirac quantization condition allows states having half
the electric charge of the W bosons which now have $n_e=2$. 
Though this possibility is
not realized in the pure gauge theory, for $N_f >0$ it will
be, the elementary quarks $(\pm 1,0)$ giving an example. Also, $a$ is now defined by
$\langle\phi\rangle = a\sigma ^3$.
Thus with respect to the ``old" conventions of [\SW,\FB] we multiply $n_e$ by $2$ and
divide $a$ by $2$. Then  \di\ is still the correct mass formula.
The Picard-Fuchs equation is
$$
\left[(u^2-1){{\rm d}^2\over {\rm d}u^2} + {1\over 4}\right] 
\pmatrix{a_D\cr a\cr}(u) = 0 \ ,
\eqn\dv$$
which is equivalent to a hypergeometric differential equation with
parameters $a=b=-{1\over 2}$,\penalty -1000 $c=0$.
These parameters are not generic
but imply that the solutions are of the logarithmic type [\ERD] which
in turn ensures that one gets $SL(2,$\Z $)$ valued monodromy matrices.
The requirement that $\ad (u)$ vanishes at $u=1$, so that,
according to \di, the magnetic
monopole with $(n_e,n_m)=(0,1)$ has vanishing mass at this point,
dictates the choice for $\ad (u)$. Similarly, the asymptotics $a(u)\sim
\sqrt{u/2}$ as $u\to\infty$ determines
the choice for $a(u)$. One has
$$\eqalign{
\ad(u)&=i\,
{u-1\over 2}\, F\left({1\over 2}\raise 2pt\hbox{,}{1\over 2}\raise 2pt\hbox{,}2;{1-u\over 2}\right)\cr
a(u)&=\left({u+1\over 2}\right)^{\scriptstyle 1\over \scriptstyle 2}\,
F\left(-{1\over 2}\raise 2pt\hbox{,}{1\over 2}\raise 2pt\hbox{,}1;{2\over u+1}\right)
\ ,\cr }
\eqn\dvi$$
where $F$ is the standard hypergeometric function, see e.g. [\ERD].
Here and in the following, the argument of a complex
number always runs from $-\pi$ to $\pi$. 
Thus, $a_D(u)$ has a cut on the real line from 
$-\infty$ to $-1$, while $a(u)$ has two cuts, both on the real line, one from 
$-\infty$ to $-1$ and another from $-1$ to $1$, see Fig. 1.
\vskip 3.mm
\fig{The branch cuts of $a_D(u)$ (left) and of $a(u)$ (right).}{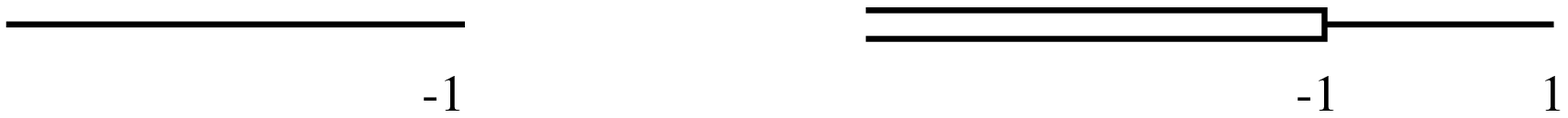}{12cm}
\figlabel\figi
\vskip 2.mm
The asymptotic behaviour of the functions $\ad(u)$ and $a(u)$ as $u\to \infty,1,-1$
is [\FB]:
$$\left.\eqalign{
\ad(u)&\simeq {i\over \pi} \sqrt{2u}\bigl[ \log u +3 \log 2-2\bigr] \cr
a(u)&\simeq {1\over 2}\sqrt{2u} } \quad \right\} \quad {\rm as\ } u\to\infty$$
$$\left.\eqalign{
\ad(u)&\simeq i\,{u-1\over 2}\cr
a(u)&\simeq {2\over \pi}-{1\over 2\pi}{u-1\over 2} \log{u-1\over 2}+
{1\over 2\pi}{u-1\over 2}\bigl( -1+4\log 2\bigr) } \quad 
\right\} \quad {\rm as\ } u\to 1$$
$$\left.\eqalign{
\ad(u)&\simeq  {i\over \pi}\left[  -{u+1\over 2} \log{u+1\over 2} + {u+1\over 2}
\bigl( 1+4\log 2\bigr) -4\right]\cr
a(u)&\simeq  {i\over 2\pi}\left[ \e {u+1\over 2} \log{u+1\over 2} + {u+1\over 2}
\Bigl( -i\pi -\e\bigl(1+4\log 2\bigr)\Bigr) +4\e\right]
 } \quad  \right\} \quad {\rm as\ } u\to -1 \ ,
\eqn\dviia$$
where $\epsilon$ is the sign of $\IM u$.
The monodromy matrices around the singular points $u=-1,1,\infty$ can
then be 
read off from the different asymptotics. One has
$$M_{\infty}=\pmatrix{-1 &\hfill 4\cr\hfill 0 &-1\cr}\ , \ 
M_1=\pmatrix{\hfill 1 &0\cr -1 &1\cr}\ , \
M_{-1}=\pmatrix{-1 &4\cr -1 &3\cr}\ , \
M_{-1}'=\pmatrix{\hfill 3 &\hfill 4\cr -1 &-1\cr}\ ,
\eqn\dviii$$
where $M_{-1}$ is to be used if the monodromy around $u=-1$ is computed
with a basepoint in the lower half $u$-plane and $M_{-1}'$ if the
basepoint is in the upper half $u$-plane. This distinction comes about
since the neighbourhoods of  $u=-1$ in the upper and lower half
planes are separated by the branch cuts.
This singularity structure corresponds to a massless
magnetic monopole $(0,1)$ at $u=1$ and to a massless dyon, alternatively
described as $(2,1)$ or $(-2,1)$, at $u=-1$.

The curve of marginal stability ${\cal C}_0$ was studied in
[\FAY ,\ARG ], see Fig. 2.
Along this curve, $\ad /a$ 
takes all the values in the interval $[-2,2]$, with 
$a_D / a$ increasing monotonically from $-2$ to $+2$ as one goes
along the curve clockwise from $u=-1+i\epsilon$ to $u=-1-i\epsilon$,
with $a_D / a=0$ at $u=0$.

\vskip 2.mm
\fig{The curve \C\ in the complex $u$-plane
where ${\displaystyle a_D\over\displaystyle a}\in \R$.}{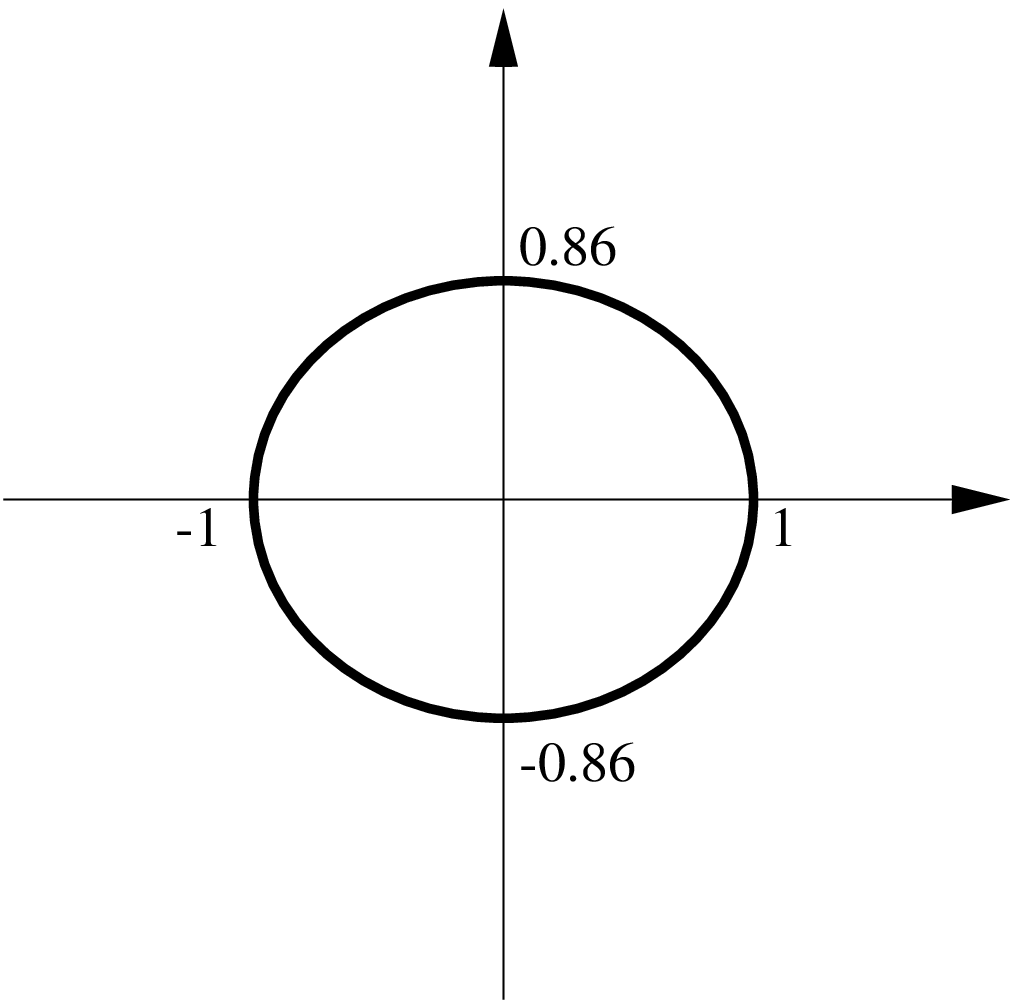}{6cm}
\figlabel\figii
\vskip 2.mm

The mass formula \di\ implies the following important property: 
a BPS state with\break
$\displaystyle {n_e \over n_m}\equiv r \in [-2,2]$ will
become massles at the point $u^*\in {\cal C}_0$ where 
$\displaystyle {a_D \over
a}(u^*)=r$. On the other hand, the only singularities at finite $u$ on
the moduli space are at $u=\pm 1$, and hence the only BPS states ever
getting massless are the magnetic monopole and the dyon.
This constraint was extensively used in [\FB] in the derivation
of the $N_f =0$ spectra. In the $N_f > 0$ case the same constraint
will also prove to be very useful.

Let us finally mention that for $N_f=0$, with the present normalisation, the
weak-coupling (semi-classical) spectrum consists of the dyons $(2n,1),\ n\in \Z$ as
well as the W-bosons $(2,0)$. The strong-coupling spectrum contains the monopole
$(0,1)$ and the dyon $(\pm2,1)$ only [\FB].

\section{Global symmetries}

Let us recall the field content of the theories we consider.
In addition to the $N=2$ vector multiplet containing the
gluons, the two gluinos $(\lambda ,\psi )$ and the Higgs
scalar $\phi $, we have $N_f$ hypermultiplets 
$(q^f ,\chi ^f ,\tilde q_f ,\tilde\chi _f)$ where
$(q^f$, $\tilde q_f)$ are complex scalars and $(\chi ^f$, $\tilde\chi _f)$
are Weyl spinors. $f$ is the flavour index running from $1$
to $N_f$ and $(q, \chi )$ and $(\tilde q,\tilde\chi )$ transform
in complex conjugate representations of the gauge group.
For $SU(2)$ these are actually equivalent representations.

Because $N=2$ supersymmetry is not broken
in this theory, we have an $SU(2)_R$ quantum
symmetry for which the gluinos and the scalars $(q,{\tilde q}^{\dagger})$
are doublets. This global exact quantum
symmetry will play no particular r\^ole
in the following.
At the classical level,
since the bare masses of the quarks are zero, we also have
an $U(1)_R$ symmetry under which the gluinos have charge 1,
the Weyl spinors in the hypermultiplets charge $-1$, and
the Higgs scalar charge 2. This symmetry 
acts on the Higgs scalar and thus potentially on the
moduli space. It will be of utmost importance in our analysis.
Finally, when $N_f >0$, which is the case on which we focus by now, 
we have the flavour symmetry which is exceptionally
$O(2N_f)$ here, thanks to the pseudo-reality of $SU(2)$ representations.
The $U(1)_R$ and $O(2N_f)$ symmetries
are anomalous. Only a subgroup
\Z $_{2(4-N_f)}$ of $U(1)_R$ survives quantum mechanically.
Moreover, the parity $\rho$ in $O(2N_f)$ is also anomalous, and 
the quantum flavour symmetry is really $SO(2N_f)$, or more
precisely $Spin(2N_f)$, as the theory contains spinor
multiplets of $SO(2N_f)$ (typically, semi-classical
reasoning shows that this is the case for the particles with unit
magnetic charge [\SWII ,\SET ,\GAU ]).
This parity $\rho $, which
exchanges $(q, \chi )$ and $(\tilde q, \tilde\chi )$ for one flavor
of quark and leaves the other fields unchanged, when
composed with the $U(1)_R$ symmetry, gives rise to an improved exact
\Z $_{4(4-N_f)}$ quantum symmetry, whose square is nothing but
the \Z $_{2(4-N_f)}$ coming from $U(1)_R$ alone.

\subsection{\Z $_4$ symmetry at fixed $u$}

At a given point $u$ of the Coulomb branch, 
this \Z $_{4(4-N_f)}$ symmetry is spontaneously broken by the
vacuum expectation value of the Higgs field down to
\Z $_4$. This \Z $_4$ changes the sign of the Higgs field and thus
of the electric and magnetic charges. This explicitly shows
that when the state $(n_e ,n_m)$ is present, then the
anti-particle $(-n_e ,-n_m)$ is also present. Thus we will
often denote these simply by $\pm (n_e ,n_m)$. However, note that,
unlike for $N_f=2$, 
for $N_f=1$ and $N_f=3$ this apparently innocent
charge conjugation operation does include a $\rho $ transformation.
For $N_f=1$, this simply changes the sign of the $U(1)$ flavour
charge $S$.
For $N_f=3$,
$\rho $ acts as the outer automorphism of the corresponding
Lie algebra ($D_3$ in the Cartan classification) and 
will change the $Spin(6)$ representation under which the 
$(n_e ,n_m)$ state transforms into its complex conjugate. 
For the spinor representations this amounts to changing the chirality.
For $N_f=2$, particles and anti-particles occur in the same representations 
of $Spin(4)$.

\subsection{\Z $_{4-N_f}$ symmetry on the Coulomb branch}

The spontaneous breakdown of the \Z $_{4(4-N_f)}$ to the \Z $_4$ symmetry
is at the origin of the \Z $_{4-N_f}$
symmetry on the Coulomb branch of the moduli space. This is \Z $_3$ 
for $N_f =1$ or \Z $_2$  for $N_f =2$, where the broken
generators will relate {\it different} points on the Coulomb branch
where physically equivalent theories 
must lie.  For $N_f=3$,
no such symmetry exists. 
The generator of this \Z $_3$ (for $N_f=1$) or \Z $_2$ (for
$N_f=2$) symmetry always contains a $\rho $ transformation and thus
in particular will act on the flavour
quantum numbers of the states. For $N_f=1$ it changes the sign
of the $U(1)$ flavour charge. For $N_f=2$, the flavor group is
$Spin(4)=SU(2)\times SU(2)$ and $\rho $ has the effect of
interchanging the two $SU(2)$ factor (this amounts to
changing the chirality for the spinor representations).
Note also that in all cases the generator of the symmetry acting on the
Coulomb branch shifts the $\theta$ angle by $\pi $ and thus should
shift the
electric charge $n_e$ of a state $(n_e, n_m)$ by $n_m$.
This will be explicitly checked in the following.

\subsection{Relations between the quantum numbers}

We list below some constraints on the
representation of the flavour group $Spin (2N_f)$ a BPS state
$(n_e ,n_m)$ can carry. These constraints come essentially
from the semi-classical approach which allows one to obtain
the general form of the wave function for a $(n_e ,n_m)$ state,
valid in the regime where the field theory is weakly coupled.
Semi-classical reasoning is likely to give the correct answer for
the quantum numbers of the states (though not for others quantitative
features like the physical mass). We will explain in Section
3 how this is corrected by non-perturbative effects.
For $N_f =1$, the flavour group is $SO(2)=U(1)$, and the 
abelian charge $S$ under this $U(1)$ must be such that
$2S=n_m-2n_e$ mod 4. Note that this charge $S$ is not the abelian charge
appearing in the central charge $Z$ of the susy algebra 
when the bare masses of
the hypermultiplets are non vanishing [\SWII].
For $N_f =2$, the flavour group is $Spin(4)=SU(2)\times SU(2)$
and the irreducible
representations can be labeled like $(2s_1 +1,2s_2 +1)$ where
$s_1$ and $s_2$ are the spins of the corresponding $SU(2)$ representations.
Then $s_1$ must be half  integer if $n_e +n_m$ is odd, 
and must be  integer if $n_e +n_m$ is even. The same is true for 
$s_2$ with $n_e$ replacing $n_e +n_m$ (note that we could interchange
the r\^oles of $s_1$ and $s_2$; the convention chosen here and below
always correspond to the conventions of [\SWII]).
For $N_f =3$ the flavour group is $Spin(6)=SU(4)$. If $2n_e +n_m$ is
odd, we have a faithful representation of $SU(4)$. The cases 
$2n_e +n_m =1$ mod 4 and
$2n_e +n_m =3$ mod 4
correspond to complex conjugate representations. We will choose
that the monopole $(0,1)$ is in the defining representation $\bf 4$
of $SU(4)$.
If $2n_e +n_m =2$ mod 4 we have a faithful representation
of $SO(6)$, and finally if $2n_e +n_m$ is a multiple of four
we have a faithful representation of $SU(4)/$\Z $_4$, where here
\Z $_4$ is the center of $SU(4)$, or the trivial representation.

We can be more precise for the states having unit magnetic
charge. These are always in spinorial representations.
A state $(n_e,+1)$ has $S=1/2$ or $S=-1/2$ (for $N_f=1$),
it is in $(\bf 2,\bf 1)$ or in $(\bf 1,\bf 2)$ (for $N_f=2$), and it is in
$\bf 4$ or in $\overline{\bf 4}$  (for $N_f=3$), where in each case the first entry
corresponds to $n_e$ even and the second entry to $n_e$ odd. 
Concerning the $S$ charge of the quarks for $N_f=1$, note that half of the states
$(+1,0)$ have $S=+1$, and the other half have $S=-1$.

All these constraints are perfectly consistent with the former analysis
of the action of the global symmetries. Here we check this 
only for $N_f=3$, and leave the other cases to the reader.
Note that $2n_e+n_m=1$ mod 4 is equivalent to
$-2n_e-n_m=3$ mod 4, and thus that charge conjugation amounts to
complex conjugating the representations in this case, as it should.
The cases $2n_e+n_m=0$ mod 4 and $2n_e+n_m=2$ mod 4 are left
invariant by charge conjugation; this again is consistent since
a faithful representation of $SO(6)$ or $SU(4)/$\Z $_4$ gives
another faithful representation of the corresponding groups by
complex conjugation. Let us now check the action of the
generator of the \Z $_3$ symmetry.  
It indeed acts as charge conjugation as it contains a $\rho $
transformation, since $2(n_e+n_m)+n_m=-2n_e-n_m$ mod 4.

\section{$\Z _{4-N_f}$ symmetry and BPS states}

In this subsection, we will rephrase and sharpen some of the arguments
already mentioned in Section 4.1 of [\FB].
The supersymmetry charges, which we will denote by
$Q_{\alpha ,I}$ in chiral notation ($1\leq I\leq 2$ for $N=2$
supersymmetry), have charge one under the \Z $_{4(4-N_f)}$
symmetry. The relevant piece of the supersymmetry algebra
for our purposes is
$$\{ Q_{\alpha ,I},Q_{\beta ,J}\}=2\, \epsilon _{\alpha\beta }
\, \epsilon _{IJ}\, Z \ .
\eqn\dix $$
This shows that the central charge $Z=an_e -a_D n_m$ has charge
two. $Z$ can be charged under
a global discrete symmetry group, since such a group
is not generated by infinitesimal generators belonging to the
algebra. The transformation law of $Z$ under the action of the
generator of the \Z $_{4-N_f}$ symmetry acting on the moduli
space is then
$$Z\longrightarrow Z'=e^{\pm 
{\scriptstyle i\pi\over\scriptstyle 4-N_{\scriptscriptstyle f}}}\, Z \ .
\eqn\dx $$
Thus we have $\vert Z'\vert =\vert Z\vert$. Moreover, since 
the \Z $_{4-N_f}$ symmetry of course  maps states of the same
physical mass into each other ($m'=m$), we see that a BPS state
will be mapped into another BPS state: $m=\sqrt{2}\,\vert Z\vert$
implies $m'=\sqrt{2}\,\vert Z'\vert$. This fact could also be
understood in terms of conservation of the number of degrees of freedom,
as a BPS state lies in a short representation
of the supersymmetry algebra unlike the other states with
$m>\sqrt{2}\,\vert Z\vert$.
Moreover, if $u$ and $u'$ denote two points related by \Z $_{4-N_f}$,
and if there exists
a matrix $G\in SL(2,$\Z$)$ and a phase $e^{i\omega}$ determined
for instance by \dx\ such that
$$\pmatrix{a_D\cr a\cr}(u')=e^{i\omega}\, G
\pmatrix{a_D\cr a\cr}(u) \ , \,\pmatrix{n_e'\cr n_m'\cr}=G\pmatrix{
n_e\cr n_m\cr} \ ,
\eqn\dxi $$
the relation $\vert Z'\vert =\vert Z\vert $ will be obviously satisfied.
We will show in the following that relation \dxi\ is indeed realized.

To summarize, the existence of
a BPS state $p=(n_e ,n_m)$ at $u$ implies the existence
of another BPS state $p'=(n'_e ,n'_m)$
 at $u'$, $u$ and $u'$ being related by a
global symmetry acting on the moduli space, with the
following properties: they have the same mass, their electric and magnetic
quantum numbers are related as in \dxi , and the representations
of the flavour group in which they lie, though
they may be different as explained in
Section 2.2, have the same dimension.

\section{A note on some mild assumptions made in this work and some
of their consequences}

On the Coulomb branch, the coordinate $u$ labels physically inequivalent
theories. The operators $\cal O$ corresponding to the observables of the
theory at $u$ may thus be labeled by $u$, and act in a Hilbert
space ${\cal H}_u$. It is natural to suppose that, at least
locally, ${\cal H}_u$ does not depends on $u$ and that the eigenvalues
and eigenfunctions of the operators
${\cal O}_u$ vary continuously with $u$.
Thus, the mass
$m(u)$ of a BPS state, as given in \di , must 
be a continuous function of $u$ as one moves in the
Coulomb branch. This should also be the case for
the electric and magnetic charges, but as these are integers,
they must be constant. For the same reason,
the Witten index or its generalizations,
used to show that supersymmetry
is not broken in the models we consider, must be constant.

This picture cannot be maintained globally. The reason
for this is that there exist curves of marginal stability, separating
strong and weak-coupling regions,
allowing the ``decays'' of usually stable BPS states. In
[\FB ], it was shown that such decays {\it do} happen in the
pure gauge theory, and one of the most important results of the present
paper will be to explicitly demonstrate that this is also the case for the 
$1\leq N_f \leq 3$ theories. Thus, at least in the
BPS sector of the Hilbert space, there must be some kind of
discontinuity on this curve. The transition between the two sides
of the curve may be understood as follows. The theories which
lie just on the curve should have physically equivalent sectors
in their Hilbert space. The predicted ``decay'' patterns on the curve
are nothing but the rules for this identification. If this is
correct, the quantum numbers of the states related by such a decay
should be compatible, and we will see that this is always the case.
Once the identification is made, one can forget about the different
sectors and only work in a reduced Hilbert space
which corresponds to one given sector. Now, when we leave the
curve, moving into the strong-coupling region, we are left with only
the reduced Hilbert space! Typically, the BPS sector of this
reduced Hilbert space will contain only the states which are responsible
for the singularities on the Coulomb branch.
Note that it is not clear whether the Hilbert space contains
other sectors than the BPS sector, and if this hypothetic new
sectors can undergo discontinuities.

Another interesting feature of the theories under study
is the following. 
If a spontaneously broken global symmetry relates two points
$u$ and $u'$ which can be linked by a path which does not cross
any discontinuity curve, then the {\it broken } symmetry
predicts the existence of certain states in a theory at {\it fixed} $u$.
We will show that, in the strong-coupling 
region, this implies that the BPS states
come in multiplets of the broken \Z $_{4(4-N_f)}$ symmetry!
In the $N_f =0$ case we had the \ZZ\ pairs of [\FB ],
and here we will have \Z $_3$ pairs for $N_f =1$ and
\ZZ\ pairs for $N_f =2$.

\chapter{One flavour of quarks}

\section{Global analytic structure, and \Z $_3$ symmetry}

We begin by discussing the case of a single massless hypermultiplet.
We present here all the arguments in quite some detail. When dealing
with $N_f=2$ and $N_f=3$ in the following sections, we will mainly focus
on the new features and try to avoid repetitions.

As was argued in [\SWII ],
for $N_f=1$ one has three singularities at finite $u$,
related by the $\Z_3$ symmetry, and due to dyons becoming massless.  
We will choose the scale and orientation
on the $u$-plane so that they are located
at the points $u_1=e^{i\pi /3}$, $u_2=-1$ and $u_3=e^{-i\pi /3}$.
Suppose that the singularity at $u_3$ is produced by a state 
$(n_e,n_m)$. The associated monodromy matrix is given by \div\ with
$d=1$ (here the flavour group is abelian and its irreducible representations
are all of dimension 1). According to the discussion in Section 2.2,
because of the \Z $_3$ symmetry the states becoming massless at 
$u_1$ and $u_2$ will be respectively $(n_e-n_m,n_m)$ and
$(n_e-2n_m,n_m)$. A general consistency condition for the
monodromy group is actually
$$M_{(n_e-2n_m,n_m),1}M_{(n_e-n_m,n_m),1}M_{(n_e,n_m),1}=M_{\infty}
=M_{(n_e-n_m,n_m),1}M_{(n_e,n_m),1}M_{(n_e+n_m,n_m),1},
\eqn\ti $$
where $M_{\infty}$ is given by \diii . This implies $n_m=\pm 1$.
The electric charge is left arbitrary,
reflecting the possibility of ``democracy'' transformations, see
[\FB ], and we choose $n_e=0$.
This shows that the ``minimal'' choices of [\SWII ] are
actually the only possible one.

Let us now determine  the analytic structure
of the $SL(2,$\Z $)$ bundle $E_1$.
The corresponding elliptic curve  is [\SWII]
$$y^2=x^3-ux^2-{\Lambda_1^6\over 64} \ ,
\eqn\tii$$
and  ${\rm d}\adu/ {\rm d} u$ and ${\rm d}\au/ {\rm d} u$
are given by its period integrals. 
Then  $\adu$ and $\au$ are given by the integrals 
over appropriate contours $\gamma_i$
of a one-form $\tilde\lambda$, where  $\tilde\lambda$ is obtained by
integrating
${\displaystyle{\sqrt{2}\over 8\pi} {{\rm d}x \over y}}$ with respect to $u$ 
(up to an exact
one-form on  the elliptic curve). This yields the
integrals
$${\sqrt{2}\over 8\pi}  \oint_{\gamma_i} {\rm d}x 
{2 u-3x\over \sqrt{x^3- u x^2 -\Lambda_1^6/64} }
\eqn\tiii$$
for $\adu$ and $\au$. The most convenient choice for the mass scale,
leading to singularities on $|u|=1$,
is such that
$$\Lambda_1^6={256\over 27}.
\eqn\tiv$$

It is straightforward to show from \tiii, and it was established in [\IY] that
$\adu$ and $\au$ satisfy the differential equation\foot
{It comes from the Picard-Fuchs equation 
satisfied by the period integrals which are the derivatives of
$\adu$ and $\au$ with respect to $u$.} 
$$\left[ \left( u^3+1\right) {{\rm d}^2\over {\rm d} u^2 } -{u\over 4}\right]
\pmatrix{ \adu\cr \au\cr}(u) = 0 \  .
\eqn\tv$$
This is not a hypergeometric equation, but it can be transformed into one if we change
variables [\IY] to
$$v=- u^3 \ .
\eqn\tvi$$
Then
$$\left[ v(1-v) {{\rm d}^2\over {\rm d} v^2 } + {2\over 3}
(1-v) {{\rm d}\over {\rm d} v }  - {1\over 36}\right]
\pmatrix{ \adu\cr \au\cr} = 0 \ .
\eqn\tvii$$
This is a hypergeometric differential equation 
with $a=b=-{1\over 6}$ and $c={2\over
3}$.

\vskip 2.mm
\fig{Definition of the regions $A,\, B^+,\, B^-,\, C$ and positions of the
singularities $u_1,\, u_2,\, u_3$ for $N_f=1$}{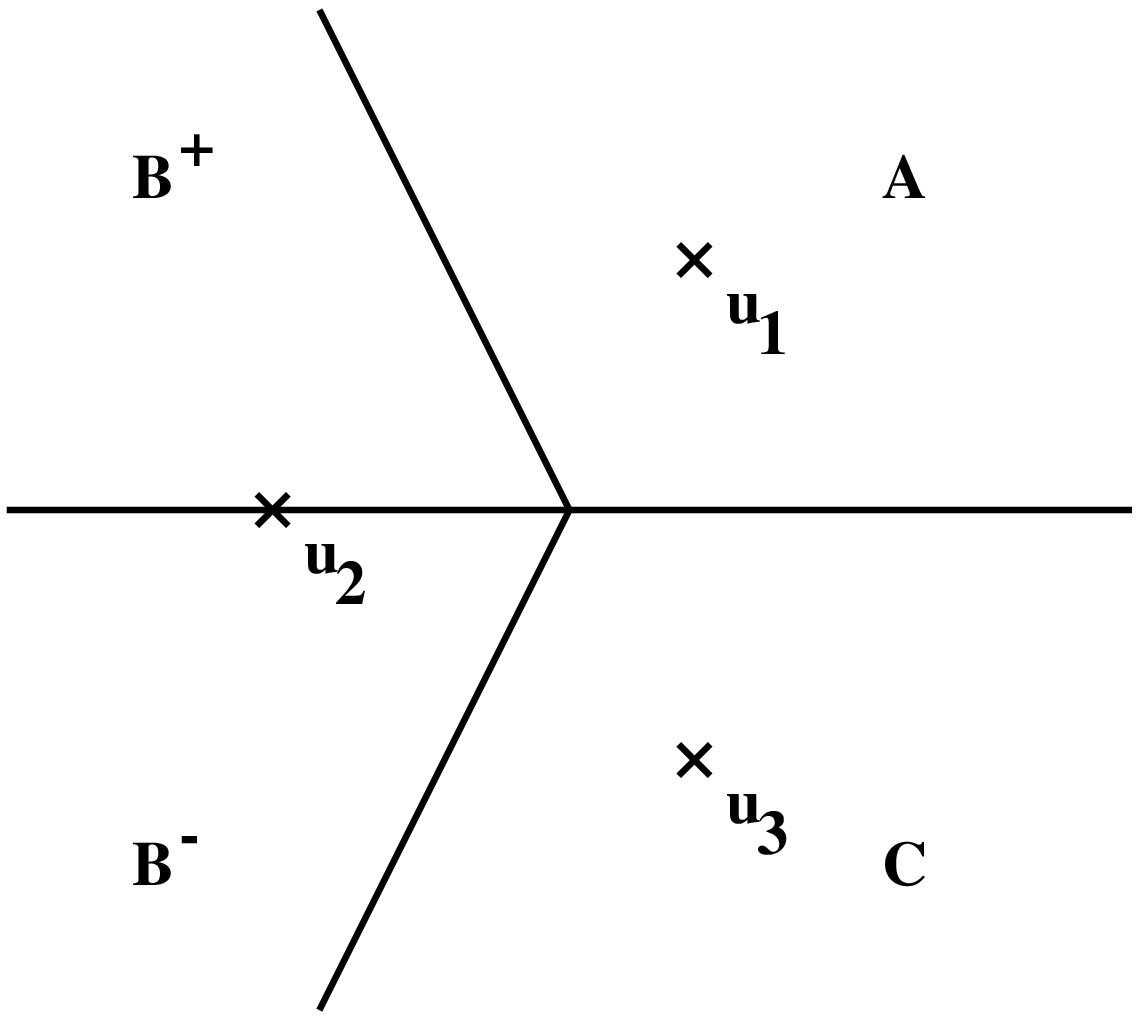}{6cm}
\figlabel\figvii
\vskip 2.mm
Note that the mapping \tvi\  is 3 to 1 (except at $u=0$)
and each of the three regions $A$, $B$
($\equiv B^+\cup B^-$) and $C$ of the $u$-plane shown in Fig. 3 are mapped onto the
full $v$-plane.  The singular points of the differential equation \tvii\ 
are $v=0, 1$ and
$\infty$, corresponding in the $u$-plane to
$$u_0=0 \ , \  u_1=e^{i\pi/3}\ , \ u_2=-1    \ , \ 
u_3 =e^{-i\pi/3}\ , \ u_\infty=\infty \ .
\eqn\tviii$$
The borders between the regions $A,\, B$ and $C$ are mapped onto the
negative real $v$-axis. Any solution of eq. \tvii\ on the $v$-plane will yield a solution
of \tv\ in any one of the three regions $A,\, B$ or $C$. However,
to obtain a solution on the whole $u$-plane we must choose some solution in a given
region, say in $C$, and then analytically continue it into the
neighbouring regions.
Note also that  although $v=0$ is a singular point of \tvii , $u=0$  of course
is not a  singular
point of \tv .

A convenient choice of two linearly independent solutions to the differential equation
\tvii\ are Kummer's solutions
$\tilde U_3(v)$ and $U_6(v)$ [\ERD]:
$$\eqalign{
U_6(v)&= (1-v) F\left( {5\over 6}\raise 2pt\hbox{,} {5\over 6}\raise 2pt\hbox{,}2;1-v\right) \cr
\tilde U_3(v) &= v^{1/6}F\left( -{1\over 6}\raise 2pt\hbox{,} {1\over 6}\raise 2pt\hbox{,}1;{1\over v}\right) \ . \cr }
\eqn\cvi$$
The first solution $U_6(v)$ obviously vanishes as $v\to 1$, and hence, once
appropriately normalized, is a good candidate for $\adu$. The second solution behaves
for $v\to\infty$ as
$$ v^{1/6} = (-u^3)^{1/6}= \omega(u) \sqrt{u} \ ,
\eqn\cvii$$
where $\omega(u)$ is a region-dependent phase factor: 
$\omega(u)=e^{-i\pi/2}$ for $u\in B^+$, 
$\omega(u) = e^{-i\pi/6}$ for $u\in A$, 
$\omega(u)=e^{i\pi/6}$ for $u\in C$ and
$\omega(u)=e^{i\pi/2}$ for $u\in B^-$.
Once correctly normalized, $\tilde U_3$ is thus a good candidate for $\au$.  The 
asymptotics
\cvii\ are discontinuous, but as explained above, one should take the solution in
one region  and then analytically continue it to the other regions. 
The correctly normalized function $\au$ then is
$$\au(u)={1\over 2}\sqrt{2u}\, F\left( -{1\over 6}\raise 2pt\hbox{,} {1\over 6}\raise 2pt\hbox{,}1;-{1\over u^3}\right)
\ .
\eqn\cviii$$
This function has one branch cut along the negative real $u$-axis from the
square-root, as well as three other cuts extending from the origin of
the $u$-plane to $u_1,\, u_2$ and $u_3$ due to the hypergeometric function, see Fig. 4.
The expression \cviii\ for $\au(u)$ obviously has the correct asymptotics $\sim 
{1\over 2}\sqrt{2u}$  everywhere as $u\to\infty$.

\vskip 2.mm
\fig{The cuts of $\au(u)$}{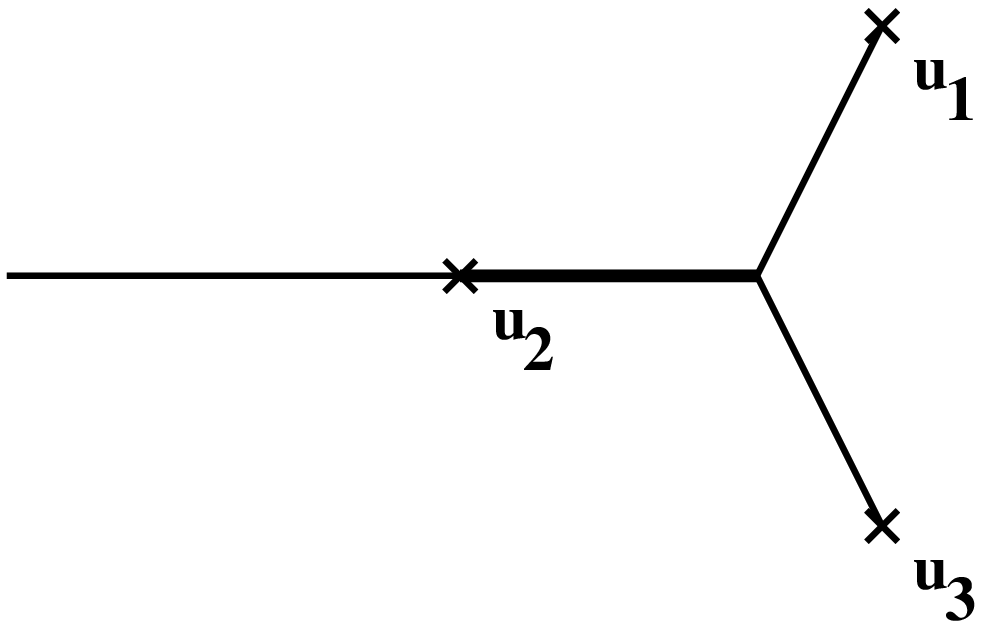}{5cm}
\figlabel\figviii
\vskip 2.mm

Next, we determine $\adu(u)$. It should be proportional to $U_6(v)$ as given in \cvi\
in the region $A$ or $B$ or $C$ in which the magnetic monopole becomes
massless. Choosing this region is a matter of convention, just as for $N_f=0$ in [\SW]
it was chosen to have the massless monopole at $u=1$ (rather than at $u=-1$). We
choose this region  to be $C$ with the  monopole becoming massless at 
$u_3=e^{-i\pi/3}$. The normalisation of $\adu$ then is determined
from  the desired asymptotics \dii . If we define for all $u$ the function
$$f_D(u) ={\sqrt{2}\over 12} \left( u^3 +1\right)
           F\left( {5\over 6}\raise 2pt\hbox{,} {5\over 6}\raise 2pt\hbox{,}2;1 + u^3 \right)
\eqn\cviiia$$
then for $u\in C$ we have
$$\adu(u)= e^{-2i\pi/3} f_D(u)\ , \quad u\in C \ .
\eqn\cix$$
Indeed, as $u\to\infty$ in region C, the asymptotics of this expression
 is
$${3i\over 4\pi} \sqrt{2u} 
\left[ {1\over 3} \log (-u^3) +{4\over 3}\log 2 -2 +\log 3 \right]
= {3i\over 4\pi} \sqrt{2u} 
\left[ \log u  +{4\over 3}\log 2 -2 +\log 3 +{i\pi\over 3} \right] \ ,
\eqn\cx$$
in accordance with the asymptotics derived directly from the integral \tiii\ [\IY].
%
%
%
One finds for all analytic continuations
$$\eqalign{
 u\in B^-       \ &: \quad  \adu(u) = - f_D(u)            +  \au(u)   \ , \cr
 u\in C^{\phantom -}\ &: \quad  \adu(u) = e^{-2i\pi/3} f_D(u)             \ , \cr
 u\in A^{\phantom -}\ &: \quad  \adu(u) = e^{-i\pi/3}  f_D(u) -  \au(u)   \ ,  \cr
 u\in B^+       \ &: \quad  \adu(u) = + f_D(u)            - 2\au(u)   \ . \cr
 }
\eqn\cxi$$
We recall that $\au(u)$ was given in \cviii\ for all $u$.
Since the different expressions for $\adu(u)$ are obtained by analytic continuation
through the cuts of $ F\left( {5\over 6}, {5\over 6},2;1 + {4\over 27} u^3 \right)$,
which are the borders between regions $B^+$ and $A$, $A$ and $C$, $C$ and $B^-$, the
function $\adu(u)$ so defined has no cuts at these borders. The only such cut 
at the border
between two regions is between $B^+$ and $B^-$. Moreover, in region $A$, due to the
presence of $\au(u)$ in the definition of $\adu(u)$, there is also a cut from $u=0$ to
$u=u_1$. The cuts of $\adu(u)$ are shown in Fig. 5. 
\vskip 2.mm
\fig{The cuts of $\adu(u)$}{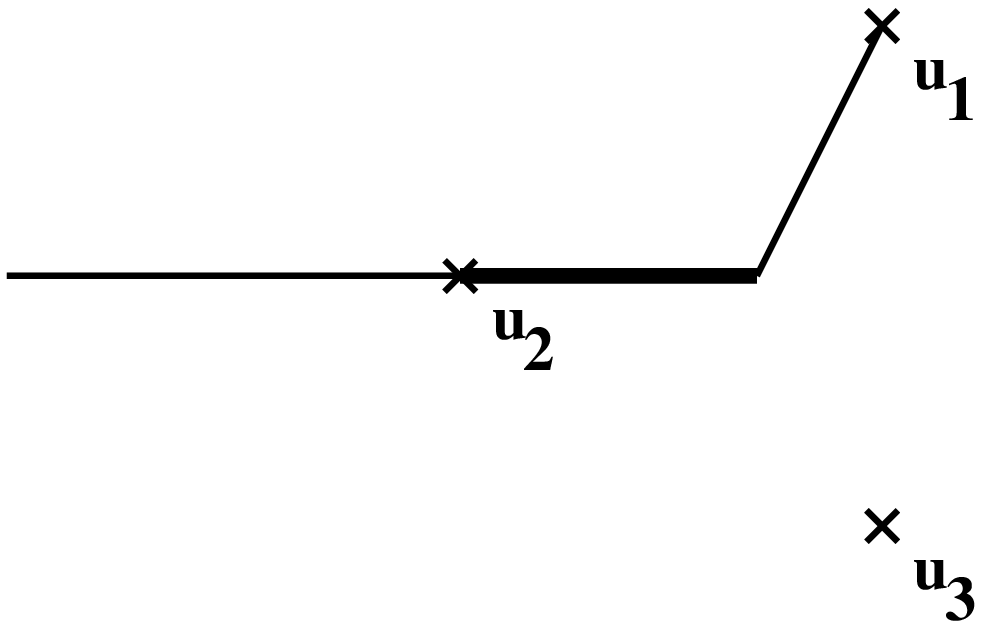}{5cm}
\figlabel\figix
\vskip 2.mm

The solutions \cxi\ reflect the $\Z_3$ symmetry on the moduli space. One  has
$$\pmatrix{ \adu\cr \au\cr } \left( e^{\pm 2i\pi/3} u\right)
= e^{\pm i\pi/3} G_{W\pm} \pmatrix{ \adu\cr \au\cr } (u) \quad , \quad 
 G_{W\pm} = \pmatrix{ 1& \mp 1\cr 0 & \hfill 1\cr} \ ,
\eqn\cxii$$
provided that $u$ is such that the path  
$t \mapsto e^{\pm 2i\pi t/3} u, \ t\in [0,1]$ does
not cross the cut on the negative real $u$-axis, see Fig. 6.
\vskip 2.mm
\fig{The matrices $e^{\pm i\pi/3} G_{W\pm}$ provide the analytic continuations
associated with the $\Z_3$ symmetry.}{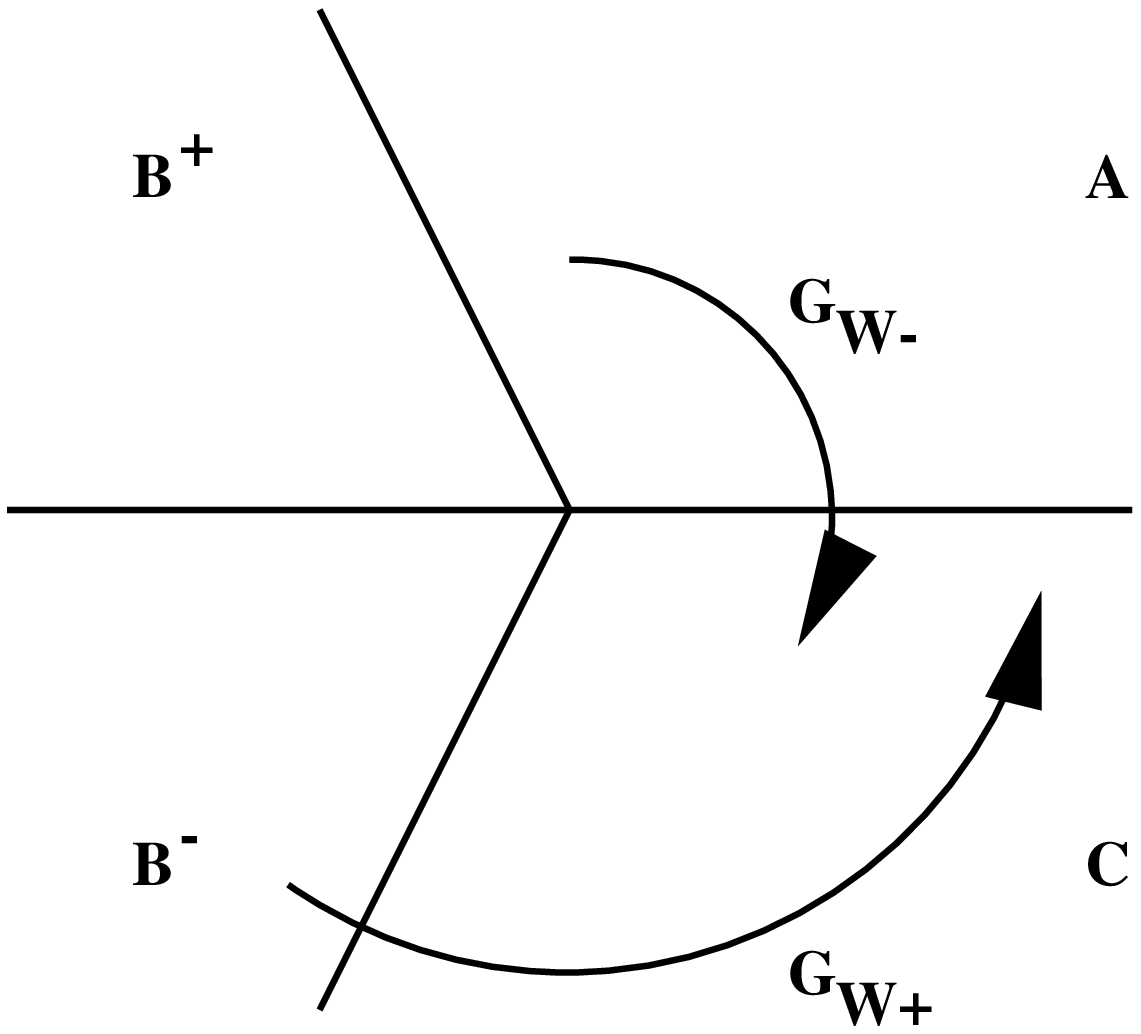}{5cm}
\figlabel\figx
\vskip 2.mm

We now proceed to give the various asymptotics and monodromy matrices for the singular
points. First, of course, $u=0$ is not a singular point, as we already noticed from the
differential equation \tv , although this is not completely manifest on our solution. 
However, for $u\to
0$, $\adu(u)$ and $\au(u)$ have a power series development of the form 
$c_0 + c_1 u + \ldots$,
{\it without} any logarithms.
The constants $c_0, c_1, \ldots$ have different phases on different sides
of the cuts going through $u=0$, as shown in Fig. 4 and 5. 
Nevertheless, if one takes the
asymptotic form at some point $u$ close to $0$ and analytically continues it
along a path surrounding $u=0$  
one obviously gets back the same series, and the monodromy is trivial, as it should. 
For the other asymptotics we find:
$$\left.\eqalign{
\au(u)&\simeq {1\over 2} \sqrt{2u} \cr
\adu(u)&\simeq {3i\over 4\pi}\sqrt{2u}\left[\log u +{4\over 3} \log 2-2+\log 3
 +{i\pi\over 3}\right] }
\quad \right\} \quad {\rm as\ } u\to\infty$$
$$\left.\eqalign{
\au(u)&\simeq {\sqrt{2}\over 8\pi }  e^{5i\pi/6} 
       \Bigg[12 e^{-2i\pi/3}  +  \bigl(u-u_1\bigr) \log \bigl(u-u_1\bigr) \cr
       &\phantom{\simeq{\sqrt{2}\over 8\pi  }  e^{5i\pi/6} }
       + \bigl(u-u_1\bigr)\left( 5-4\log 2 -2\log 3 -{i\pi\over 3}\right) \Bigg] \cr
\adu(u)&\simeq - \au(u) +  { \sqrt{2}\over 4 } e^{i\pi/3} \bigl(u-u_1\bigr) }
\quad \right\} \quad {\rm as\ } u\to u_1$$
$$\left.\eqalign{
\au(u)&\simeq  {\sqrt{2}\over 8\pi } i \e
       \Bigg[12  +  \bigl(u-u_2\bigr) \log \bigl(u-u_2\bigr) \cr
       &\phantom{\simeq{\sqrt{2}\over 8\pi  } i \e  }
       + \bigl(u-u_2\bigr)\left( 5-4\log 2 -2\log 3 +i\pi\e \right) \Bigg] \cr
\adu(u)&\simeq - {3\e+1\over 2} \au(u) +   { \sqrt{2}\over 4 } \, \e\ \bigl(u-u_2\bigr) }
\quad \right\} \quad {\rm as\ } u\to u_2$$
$$\left.\eqalign{
\au(u)&\simeq {\sqrt{2}\over 8\pi }  e^{-5i\pi/6} 
       \Bigg[12 e^{2i\pi/3}  +  \bigl(u-u_3\bigr) \log \bigl(u-u_3\bigr) \cr
       &\phantom{\simeq{\sqrt{2}\over 8\pi }  e^{-5i\pi/6} }
       + \bigl(u-u_3\bigr)\left( 5-4\log 2 -2\log 3 +{i\pi\over 3}\right) \Bigg] \cr
\adu(u)&\simeq  { \sqrt{2}\over 4 } e^{2i\pi/3} \bigl(u-u_3\bigr) }
\quad \right\} \quad {\rm as\ } u\to u_3
\eqn\cxiii$$
where $\e$ is the sign of $\IM u$. The monodromy matrices follow as
$$M_\infty=\pmatrix{-1&\hfill 3\cr\hfill 0&-1\cr}\ ,\ 
M_{u_1}=\pmatrix{\hfill 2&1\cr -1&0\cr}\ ,\ 
M_{u_3}=\pmatrix{\hfill 1&0\cr -1&1\cr}\ ,\ $$
$$M_{u_2}=\pmatrix{\hfill 0&1\cr -1&2\cr}\ ,\ 
M_{u_2}'=\pmatrix{\hfill 3&\hfill 4\cr -1&-1\cr}\ ,\ 
\eqn\cxiv$$
where $M_{u_2}$ is to 
be used for a basepoint in the lower half $u$-plane ($\e=-1$) and
$M_{u_2}'$ for a 
basepoint in the upper half $u$-plane ($\e=+1$). 
We have (cf. \ti ) $M_{u_1}M_{u_3}M_{u_2}=M_\infty=M_{u_2}'M_{u_1}M_{u_3}$.
As a check one
verifies from \cxiii\ that 
the particles that become massless are: the magnetic
monopole $(0,1)$ at $u=u_3$, 
the dyon $(-1,1)$ at $u=u_1$ and, at $u=u_2$, the dyon
described as $(-2,1)$ 
if $\e=+1$ or as $(1,1)$ if $\e=-1$. 

\section{The curve of marginal stability}

The curve of marginal stability \C $_1$ is defined as the set of all $u$ on the
moduli space such that $\wu(u)\equiv \adu(u)/\au(u)$ is real. We already know that this
curve has to pass through $u_1,u_2$ and $u_3$, since massless particles must occur on
this curve, and we have $\wu(u_3)=0,\ \wu(u_1)=-1$, etc. As for $N_f=0$, one can try to
determine the curve \C $_1$ analytically as follows: the Picard-Fuchs differential equation
\tv\ implies that $\wu$ satisfies
$$\{\wu,u\}=-2 {u\over u^3+1}
\eqn\cxv$$
where $\{ f,u \}=f'''/ f' -{3\over 2}\left( f''/ f' \right)^2$ denotes the
Schwarzian derivative of $f$ with respect to $u$. The inverse function $u(\wu)$ then
satisfies the differential equation
$${u'''\over u'}-{3\over 2}\left( {u''\over u'}\right)^2 = 2 (u')^2  {u\over u^3+1}
\eqn\cxvi$$
and it is enough to find the solution for a  real argument $\wu$ that satisfies the
appropriate initial conditions $u(0)=u_3$, $u(-1)=u_1$, etc. 

\vskip 2.mm
\fig{The curve of marginal stability for \smash{$N_f=1$}
passes through the three cubic roots 
$u_1, u_2, u_3$ of $-1$. It is almost a circle. The
numbers $-2, -3/2, \ldots, 1$ indicate the values taken by 
\smash{$\wu=\adu/\au$} along
this curve.}{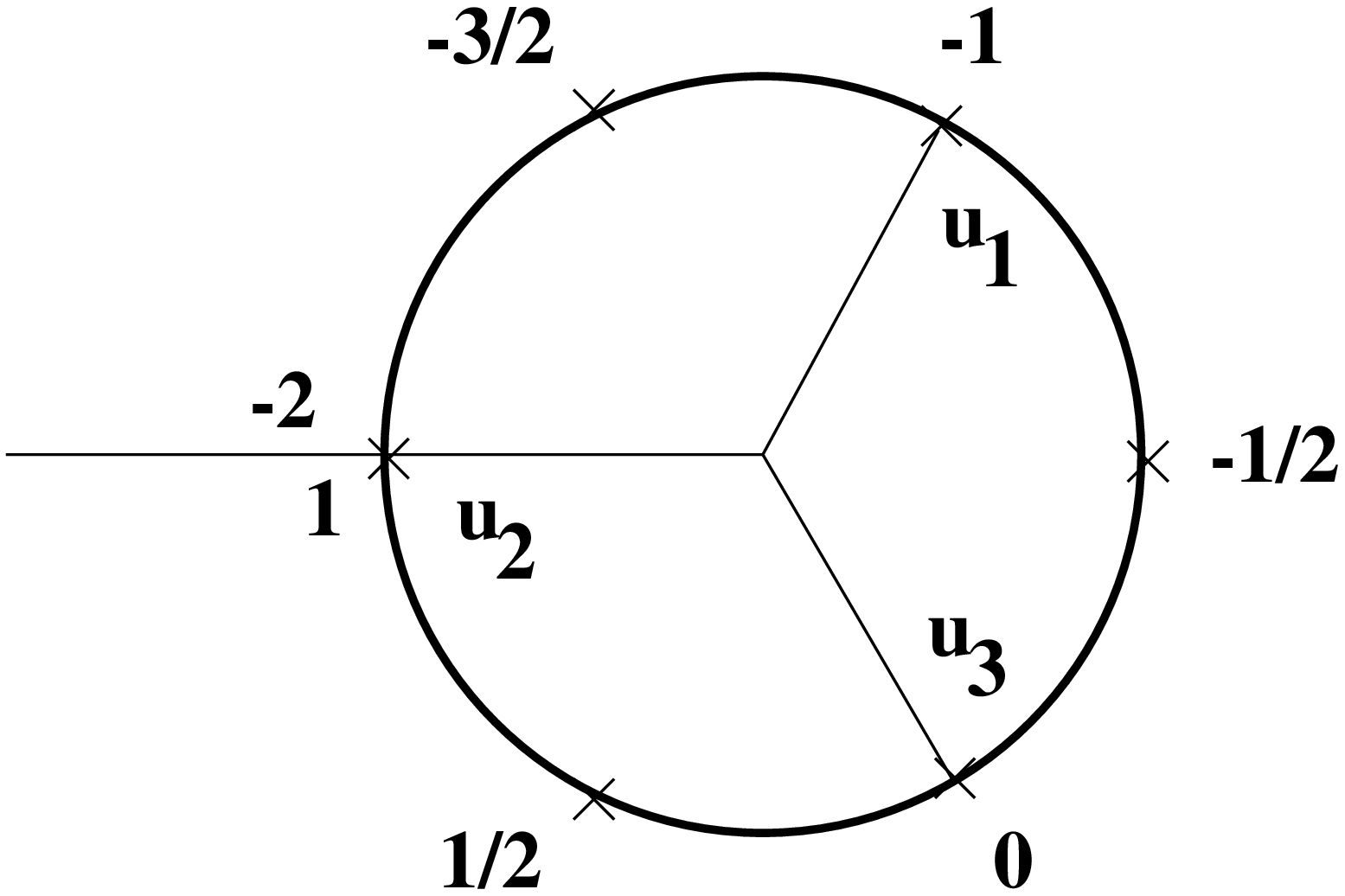}{7cm}
\figlabel\figxi
\vskip 2.mm
But again, the precise determination of the curve \C $_1$ is completely irrelevant for our
purpose. Instead, just as we did in [\FB], we have determined \C $_1$ numerically, directly
from the condition $\IM\wu(u)=0$. In any case, the $\Z_3$ symmetry \cxiv\ implies that
this curve is symmetric under rotations by ${2\pi\over 3}$ around the origin. It turns
out to be almost a circle,\foot{
The distance from the origin  is slightly larger at $u=u_1, u_2, u_3$ 
and slightly smaller at the borders between
the regions $A, B, C$.}
see Fig. 7. 

If we call \CM\ the portion of \C $_1$ between
$u_2$ and $u_3$ (see Fig. 8), \CO\ the portion between $u_3$ and $u_1$ and \CP\ the
portion between $u_1$ and $u_2$, one can show that
$$u\in {\cal C}^+ \ \Leftrightarrow\ \wu \in [-2,-1] \quad , \quad
u\in {\cal C}^0 \ \Leftrightarrow\ \wu \in [-1,0] \quad , \quad
u\in {\cal C}^- \ \Leftrightarrow\ \wu \in [0,1] 
\eqn\cxvii$$
with $\wu(u)$ increasing monotonically as $u$ goes around the curve in the clockwise
sense starting at $u_2$.

\section{The weak-coupling spectrum}

Let us first determine the 
spectrum of BPS states in the weak-coupling region \rw, i.e.
outside the curve \C. 

On the one hand we know that there are  the elementary excitations of the perturbative
spectrum, namely the W bosons $(2,0)$ and the quarks $(1,0)$. 
Note that it seems that the W are degenerate with two quarks and thus
may disintegrate as $(2,0)\rightarrow 2\times(1,0)$. However this reaction
is impossible here since the W have  flavor charge $S=0$ while the quarks
$(1,0)$ have $S=1$. Thus the W must be stable in this theory.
In the cases $N_f=2$ and $N_f=3$, such a simple argument does not
work, and we will then admit 
the existence of the W in the weak-coupling regions,
interpreting these states as bound states at threshold, as was done
in [\SWII ].
Clearly, $\pm (1,0)$ and $\pm (2,0)$ are the only stable states
of zero magnetic charge.
Now, let's turn to the
magnetically
charged states. Since the three states $(0,1),\ (-1,1)$ and $(-2,1)$ become massless at
$u_3, u_1$ and $u_2$, i.e. just on the curve \C $_1$, they must exist in both \rw\ and \rs.
The monodromy around infinity is a symmetry of the semi-classical weak-coupling
spectrum \sw, so we conclude that all states 
$\pm (n_e-3k, 1),\ n_e=0,1,2,\ k\in \Z$ must
be in \sw. But these are all the dyons $\pm (n,1),\ n\in \Z$, with unit magnetic 
charge. It
is also easy to see that one cannot have dyons with $|n_m|\ge 2$ in \sw. If
there were such a dyon $(n_e, n_m)\in$\sw\ then, again, all dyons 
$(n_e-3 k\, n_m, n_m)$ 
would be in \sw,
too. Such a dyon would become massless on \C $_1$ and lead to a singularity 
if $\displaystyle{{n_e-3k n_m\over n_m}={n_e\over n_m}}-3k\in [-2,1]$. There
is always such a $k\in\Z$. But we know that the only massless dyons have $n_m=1$. Hence
there are no dyons with $|n_m| \ge 2$ in \sw, and
$${\cal S}_W=\left\{ \pm (2,0)\ ,\ \pm (1,0)\ ,\ 
\pm (n,1)\ ,\  n\in\Z \right\} \ .
\eqn\cxx$$
Finally, recall from Sect. 2.2  that the states 
$\pm (2,0)$ have $S=0$, half of the quark states  $\pm (1,0)$ 
have $S=\pm 1$ and the other half have $S=\mp 1$. Concerning the dyons,
$\pm (2n,1)$ have $S=\pm 1/2$, and $\pm (2n+1,1)$ have
$S=\mp 1/2$.
\vskip 2.mm
\fig{The definition of the various portions of the curve and of the
strong-coupling region}{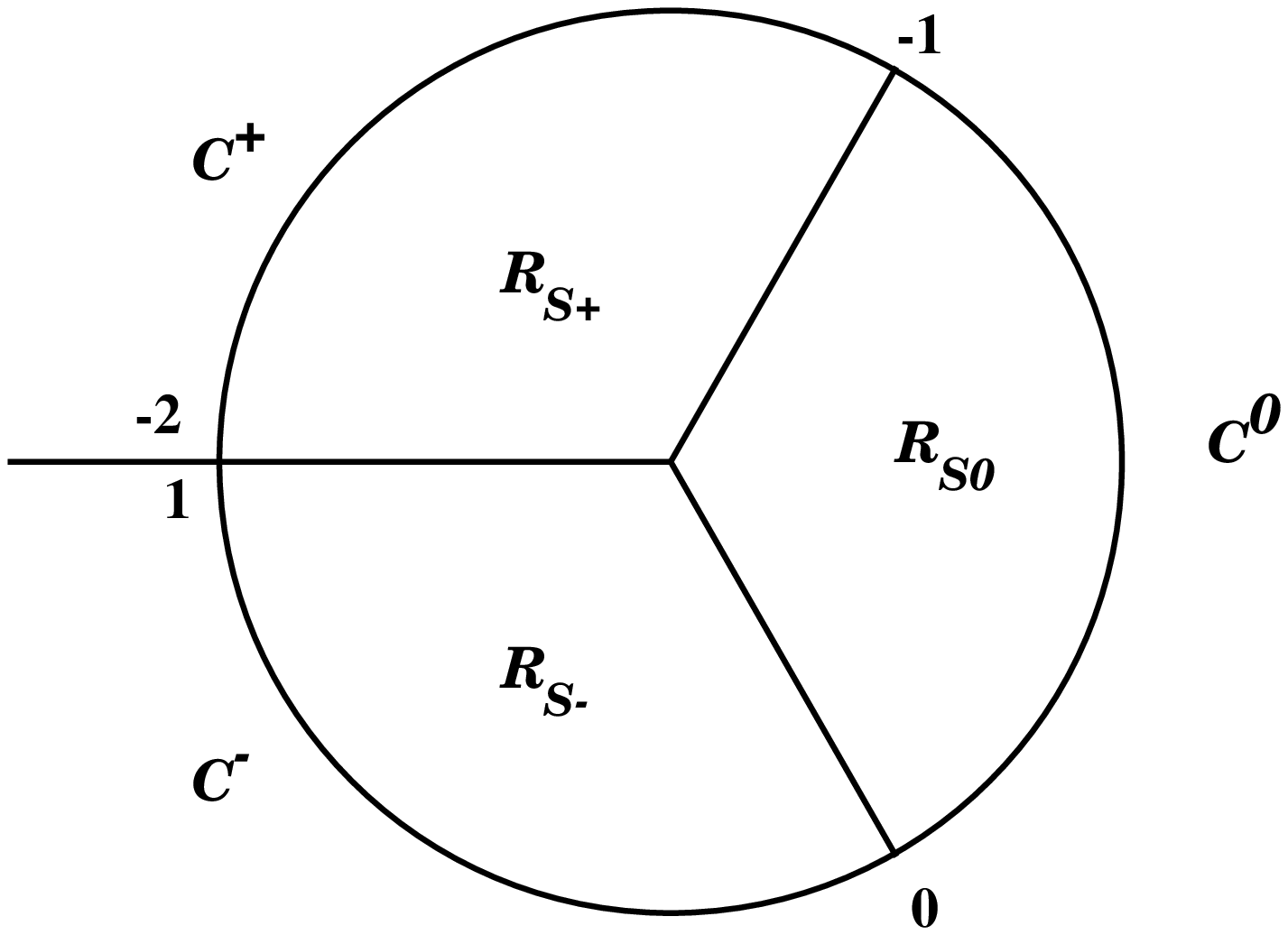}{6cm}
\figlabel\figxii
\vskip 2.mm

\section{The strong-coupling spectrum}

\subsection{Mathematical description of BPS states}

Next, we turn to the strong-coupling spectrum. As for $N_f=0$ [\FB], 
due to the cuts, see
Fig. 8, the {\it same} BPS state is described by different couples of integers in the
different regions \rsp, \rsm, \rso. We call \ssp, \ssm, \sso\ the corresponding spectra
of integers \pp, while \ss\ means the corresponding  spectrum of BPS states (unique
throughout \rs). Let's first work out these different descriptions. Suppose we have a
state described in \rso\ by \pp. Now
transport it along a path inside the
strong-coupling region \rs\ to a point $u'\in$\rsm. 
Since one does not cross the curve \C,
the state remains stable and cannot decay. The mass 
$m(u)=\sqrt{2}\,\bigl| n_e
\au(u)-n_m \adu(u)\, \bigr| $ will vary continuously, even as one crosses the cut,
since physically nothing happens there. What happens  as one crosses the cut is that
one is analytically continuing $\adu(u)$ and $\au(u)$ to some
$\tilde\adu(u'),\ \tilde\au(u')$. The latter are expressed in terms of $\adu(u'),\
\au(u')$ by using the monodromy matrix around $u_3$ as follows:
$$\pmatrix{ \tilde\adu\cr \tilde\au\cr}(u') = M_{u_3} 
\pmatrix{\adu\cr \au\cr}(u') \ .
\eqn\cxxa$$
Hence
$$m(u')=\sqrt{2}\, \Big\vert\, \eta\bigl((n_e, n_m), 
(\tilde\adu ,\tilde\au)(u') \bigr)\Big\vert
=\sqrt{2}\, \Big\vert\eta\bigl( M_{u_3}^{-1}(n_e,n_m), 
(\adu ,\au )(u') \bigr) \Big\vert
\eqn\cxxb$$
and we see that the same BPS state is described in \rsm\ by
$$\pmatrix{ n_e'\cr n_m'\cr} = M_{u_3}^{-1} \pmatrix{ n_e\cr n_m\cr}
=\pmatrix{ n_e\cr n_m+n_e \cr} \quad {\rm in} \ {\cal R}_{S-} \ .
\eqn\cxxi$$
Similarly, in \rsp\ this state is described by
$$\pmatrix{ n_e''\cr n_m''\cr} = M_{u_1} \pmatrix{ n_e\cr n_m\cr}
=\pmatrix{ 2n_e+n_m\cr -n_e \cr} \quad {\rm in} \ {\cal R}_{S+} \ .
\eqn\cxxii$$
In other words,
$${\cal S}_{S-}=  M_{u_3}^{-1}{\cal S}_{S0} \ ,\ 
{\cal S}_{S0}=  M_{u_1}^{-1}{\cal S}_{S+} \ ,\ 
{\cal S}_{S+}=  M_{u_1}M_{u_3}{\cal S}_{S-} \ 
\eqn\cxxiii$$
that is to say
$$p\equiv (n_e,n_m)\in {\cal S}_{S0} 
\quad \Longleftrightarrow \quad p\equiv \pm(n_e, n_m+n_e) \in {\cal S}_{S-}
\quad \Longleftrightarrow \quad p\equiv \pm(2n_e+n_m,-n_e) \in {\cal S}_{S+} \ ,
\eqn\cxxiiia$$
where $p$ is a unique locally constant section over $E_1$.

These results deserve some comments. First note that the sign
of the couple of integers representing $p$ in ${\cal S}_{S-}$ and
${\cal S}_{S+}$ is not always fixed
\foot{The sign is relevant here since the particle and anti-particle
carry opposite flavour charge.}
since eq. \cxxb\ does not fix the sign in \cxxi\ or \cxxii. However, in some cases we
can determine this sign.
For example, the monopole state $(0,1)$ becoming
massless at $u_3$ can be followed continuously from the weak-coupling
to the strong-coupling region if one crosses the curve of marginal stability
exactly at $u_3$ where $(0,1)$ is the only charged massless particle and
thus is stable. Thus $(0,1)$ must be described by the same couple
of integers in ${\cal S}_{S0}$ and in ${\cal S}_{S-}$.
The same reasoning applies also to the dyon $(-1,1)$ at $u_1$.
Now, the monopole $p\equiv (0,1)$ need not to be represented again by the
same electric and magnetic quantum numbers in ${\cal S}_{S+}$.
Indeed, \cxxiiia\ gives $p\equiv \pm (1,0)$ in ${\cal S}_{S+}$.
This shows that the distinction between electric
and magnetic quantum numbers in the strong-coupling region is not
clear. This is a highly non perturbative phenomenon, and is possible
due to the non-abelian monodromies. A somewhat similar phenomenon was discussed
in [\SWII ] when considering the deformation of the singularities of the
theory with non zero bare mass as the latter goes to zero.
Nevertheless, the ``monopole'' $(1,0)$ in ${\cal S}_{S+}$ should
not be confused with the elementary quark. Actually, $(1,0)$ in  ${\cal S}_{S+}$
has $\vert S\vert=1/2$ while an elementary quark has $|S|=1$. The fact that a state
$(1,0)$ can have $\vert S\vert=1/2$ does not contradict the semi-classical
constraint $2S=n_m-2n_e$ mod 4, since the strong-coupling region
is not continuously related to the semi-classical one. The only case
where the semi-classical constraints on the
quantum numbers are also valid in the
strong-coupling region is when one can go through the curve ${\cal C}_1$
without any discontinuity as explained above 
(i.e. in \sso\ and \ssm\ for the monopole,
but not in \ssp). 
Actually, when the strong-coupling region needs only be separated
in two different charts, as is the case for $N_f=2$ or 3,
this argument implies that the transition functions
of the $SL(2,$\Z $)$ bundle
are compatible with the semi-classical formula for the representations,
as we will see in Section 4 and 5. 
But for $N_f=1$ we need three charts 
and at this stage we are let with an ambiguity in the signs of
\cxxiiia\ for general $(n_e ,n_m)$.
For instance, we do not know if $(0,+1)$ in ${\cal S}_{S0}$ and
${\cal S}_{S-}$ is represented by $(+1,0)$ of $(-1,0)$
in ${\cal S}_{S+}$, and thus if the flavour charge of $(+1,0)$
in ${\cal S}_{S+}$ is $+1/2$ of $-1/2$. We will see in Section 3.5
how to lift this ambiguity.

\subsection{Determination of the strong-coupling spectrum}

Now we can determine the strong-coupling spectrum using the $\Z_3$ symmetry. 
We will work in \sso\ and then obtain ${\cal S}_{S\pm}$ simply from \cxxiii. So
let $u\in$\rso. Recall that by the general arguments of section 2.3, there must exist
a matrix $G_{W+}\in SL(2,\Z)$, satisfying \dxi\ for $u'=e^{2\pi i/3}u$. This matrix was
explicitly determined in \cxii. In section 2.3, it was shown that the existence of a
BPS state \pp\ at $u\in$\rso\ then implies the existence of a state
$G_{W+}(n_e,n_m)\in$\ssp\ at $u'\in$\rsp. This same BPS state must also exist at any
point in
\rso, but there it is described by $M_{u_1}^{-1}G_{W+}(n_e,n_m)\in$\sso, according to
\cxxiii. We define $G_{S+}=M_{u_1}^{-1}G_{W+}$ and conclude that
\pp$\in$\sso\ implies $G_{S+}$\pp$\in$\sso. In exactly the same way one can
use the $\Z_3$ symmetry with $u'=e^{-2\pi i/3}u$
and conclude that there must also be a state $M_{u_3} G_{W-}(n_e, n_m)\in$\sso.
Hence
$$\eqalign{
&G_{S+}=M_{u_1}^{-1} G_{W+} = \pmatrix{ 0&-1\cr 1&\hfill 1\cr} \ , \quad 
G_{S-}=M_{u_3} G_{W-} = \pmatrix{ \hfill 1&1\cr -1&0\cr}=
G_{S+}^{-1} \ , \quad \cr}$$
$$ G_{S+}{\cal S}_{S0}=G_{S-}{\cal S}_{S0}={\cal S}_{S0} \ , 
\eqn\cxxvii$$
i.e. \sso\ must be invariant under $G_{S+}^p$ for any integer $p$. Since
$$G_{S\pm}^3 = -{\bf 1}\ , 
\eqn\cxxviii$$
all strong-coupling states
come in $\Z_3$ triplets:
$$ \pmatrix{ n_e\cr n_m\cr} \quad
\matrix{ G_{S+} \cr \longrightarrow \cr \phantom{ G_{S+}^{-1}}\cr} \quad
\pmatrix{-n_m\cr n_e+n_m\cr} \quad
\matrix{ G_{S+} \cr \longrightarrow \cr \phantom{ G_{S+}^{-1}}\cr} \quad
\pmatrix{ -n_e-n_m\cr n_e\cr} \quad
\matrix{ G_{S+} \cr \longrightarrow \cr \phantom{ G_{S+}^{-1}}\cr} \quad
-\pmatrix{ n_e\cr n_m\cr} \ .
\eqn\cxxix$$
Remark that these \Z $_{3}$ triplets contain actually six states when one 
differentiates particles and anti-particles. They are multiplet
of the broken \Z $_{12}$ symmetry, as announced in Section 2.4.
For example, the $\Z_3$ triplet containing the magnetic monopole $(0,1)\in$\sso\
also contains $(1,0)\in$\sso, corresponding to $\pm (2,-1)\in$\ssp\ or to
$\pm (1,1)\in$\ssm\foot{We explain how to determine the correct
signs below.}, as well as $(-1,1)\in$\sso. Hence this triplet contains
precisely the BPS states that are responsible for the singularities and that 
become massless at $u_1, u_2, u_3$ on \C $_1$, as one may expect. 
Now we are going to show that this is the
only  $\Z_3$ triplet in \sso. 
Suppose  that \pp$\in$\sso. If $n_m=0$, this is either $(1,0)$ 
or $(n,0),\ n\ge 2$. In the first case it is 
part of the triplet just discussed. If it were $(n,0)$ then it would be part of a
triplet $(n,0),\ (0,n),\ (-n,n)$ in \sso, and $(0,n)$ would be an additional massless
state at $u_3$, which we know does not exist.
So suppose $n_m\ne 0$. With \pp\ also $(-n_m, n_e+n_m)$ and $(-n_e-n_m, n_e)$ 
must be in \sso. Now one of these three states will become
massless on \CO, which is the part of the curve \C $_1$ in \rso, 
if one of the three expressions
$${n_e\over n_m}\equiv r \ ,\quad 
{-n_m\over n_e+n_m} = -{1\over 1+r} \equiv \varphi_1(r) 
\ ,\quad 
{-n_e-n_m\over n_e}=-\left( 1+{1\over r}\right) \equiv \varphi_2(r) 
\eqn\cxxx$$
is in the interval $[-1,0]$. It is easy to see that the functions $\varphi_1(r)$ and
$\varphi_2(r)$ are such that either $r$ or $\varphi_1(r)$ or $\varphi_2(r)$ is always
in $[-1,0]$. Hence one or the other state of the triplet will always become massless 
somewhere on \C $_1$. This contradicts the singularity structure
and the triplet cannot be in \sso, unless it is the above-mentioned triplet
containing $(0,1)$. Thus
$$\eqalign{ 
{\cal S}_{S0}&=\left\{\pm (0,1)\ ,\ \pm (-1,1)\ ,\ \pm (1,0) \right\} \cr
{\cal S}_{S+}&=\left\{\pm (1,0)\ ,\ \pm (-1,1)\ ,\ \pm (2,-1)\right\}
=M_{u_1}{\cal S}_{S0} \cr
{\cal S}_{S-}&=\left\{\pm (0,1)\ ,\ \pm (-1,0)\ ,\ \pm (1,1) \right\}
=M_{u_3}^{-1}{\cal S}_{S0}\ . \cr}
\eqn\cxxxi$$

\section{Disintegrations}

We will now show that all states of \sw\ can consistently decay into the states of the
strong-coupling spectrum when crossing the curve of marginal stability. Suppose that
the curve is crossed somewhere on \CO, so that the available states are those of \sso.
(The other two possibilities can be discussed in exactly the same way and do work out
consistently, too).

First recall the flavour charges $S$ of these states in the strong-coupling region
\rso. Since the (anti)monopole $(0,\pm1)$ in \rso\ can be obtained
by continuous deformation from the
semi-classical one 
when crossing the curve \C $_1$ at $u_3$, it also has $S=\pm 1/2$. 
Similarly, the
(anti)dyon $(\mp 1, \pm 1)$ can cross \C $_1$ at $u_1$ and hence has $S=\mp 1/2$. To
determine the charge $S$ of $(1,0)\in$\sso, note that it is described as
$\pm(-2,1)\in$\ssp\ and as $\mp (1,1)\in$\ssm\ where 
it is massless at $u_2$ and can be
connected to the semi classical states. 
In both cases one obtains $S=\pm 1/2$ in \ssp\
and \ssm. However, as noted above, we do not know whether
the state corresponding to $(1,0)$ in \sso\
is, e.g. in ${\cal S}_{S+}$, $(2,-1)$ with $S=-1/2$
or rather $(-2,1)$ with $S=1/2$.
Hence we
only know that $(\pm 1,0)$ has $\vert S\vert =1/2$ but we cannot 
at first sight determine the sign. We
will see that {\it all } decay reactions work out consistently if we assume that
$(\pm 1,0)$ has $S=\pm 1/2$, while for the opposite choice almost all decays would 
violate the conservation of the $S$ charge. 
As we know that the decay reactions {\it must} take place, as was shown
above using arguments involving only the electric and magnetic
quantum numbers, we see that we can determine in this way 
the sign of the flavor charge in all cases where an ambiguity
remained. Below we list again the strong-coupling spectrum, now
denoting
the states as $(n_e,n_m)_S$ with the flavour charge quoted explicitly.
$$\eqalign{
{\cal S}_{S0} & = \left\{ (0,\pm 1)_{\pm 1/2}\ , \ (\mp 1,\pm 1)_{\mp 1/2} \ ,\ 
(\pm 1,0)_{\pm 1/2} \right\} \cr
{\cal S}_{S+} & =\left\{(\mp 1,0)_{\pm 1/2}\ ,\ (\mp 1,\pm 1)_{\mp 1/2}\ , 
(\mp 2,\pm 1)_{\pm 1/2}\right\} \cr
{\cal S}_{S-} & =\left\{(0,\pm 1)_{\pm 1/2}\ ,\ (\pm 1,0)_{\mp 1/2}\ ,\ (\mp 1,\mp 1)_{\pm 1/2} \right\}.
\cr }
\eqn\cxxxii$$
In this notation the weak-coupling spectrum contains the states
$(\pm 2,0)_0$, $(\pm 1,0)_{\pm 1}$, $(\pm 1,0)_{\mp 1}$,
$(\pm 2n,\pm 1)_{\pm 1/2}$ and $(\pm (2n+1),\pm 1)_{\mp
1/2}$, cf. \cxx.
All decay reactions across \CO\ must  preserve the total charges of $n_e, n_m$ and
$S$. One indeed has
$$\eqalign{
\pmatrix{ \pm 1\cr 0\cr}_{\pm 1} &\longleftrightarrow 
\pmatrix{0\cr \pm 1\cr}_{\pm 1/2} + \pmatrix{\pm 1\cr \mp 1\cr }_{\pm 1/2} \cr
\pmatrix{ \pm 1\cr 0\cr}_{\mp 1} &\longleftrightarrow 
3\times\pmatrix{0\cr \mp 1\cr}_{\mp 1/2} + 3\times\pmatrix{\mp 1\cr \pm 1\cr }_{\mp 1/2} 
+4\times \pmatrix{ \pm 1\cr 0\cr}_{\pm 1/2}\cr
\pmatrix{\pm 2\cr 0\cr}_0 &\longleftrightarrow 
2\times \pmatrix{0\cr\mp 1\cr}_{\mp 1/2}
            + 2\times \pmatrix{\mp 1\cr \pm 1\cr}_{\mp 1/2} 
+ 4\times \pmatrix{\pm 1\cr 0\cr}_{\pm 1/2} \cr
\pmatrix{\pm 2n\cr \pm 1\cr}_{\pm 1/2}  &\longleftrightarrow  
(2n-1) \times \pmatrix{0\cr\mp 1\cr}_{\mp 1/2}
            + 2n \times \pmatrix{\mp 1\cr \pm 1\cr}_{\mp 1/2} + 
4n\times \pmatrix{\pm 1\cr 0\cr}_{\pm 1/2} \cr
\pmatrix{\pm (2n+1)\cr \pm 1\cr}_{\mp 1/2}  &\longleftrightarrow 
 (2n+2) \times  \pmatrix{ 0\cr\mp 1\cr}_{\mp 1/2}
            + (2n+3) \times \pmatrix{\mp 1\cr \pm 1\cr}_{\mp 1/2}\cr
&\hphantom{\longleftrightarrow (2n+2) \times 
\pmatrix{ 0\cr\mp 1\cr}_{\mp 1/2}\qquad }\qquad 
+(4n+4)\times \pmatrix{\pm 1\cr 0\cr}_{\pm 1/2} \cr }
\eqn\cxxxiii$$
and remarkably enough all quantum numbers work out consistently.

It remains to check the mass balance, but this also works out in all cases, just as it
did for $N_f=0$ [\FB]. For example, for the first disintegration
in \cxxxiii\ one has on the l.h.s.
$m=\sqrt{2} \vert \au \vert $ while the mass on the r.h.s. is (let $r$ denote
$\adu/\au \in [-1,0]$ since one crosses the curve on \CO)
$$m= \sqrt{2}\left( \vert \adu \vert + \vert \adu + \au \vert \right)
= \sqrt{2} \vert \au \vert \left( \vert r \vert + \vert 1+r \vert \right)
= \sqrt{2} \vert \au \vert \left(  -r + 1+r  \right)
= \sqrt{2} \vert \au \vert
\eqn\cxxxiv$$
as it should.


\chapter{Two flavours of quarks}

\section{Global analytic structure, \ZZ\ symmetry and BPS spectrum}

As was argued in [\SWII ], we have for $N_f=2$ two singularities
on the Coulomb branch,
where BPS particles in a two dimensional representation of
$Spin(4)=SU(2)\times SU(2)$ become massless.
Because of the \ZZ\ symmetry, they must be located
at two points related by this symmetry; we choose the scale of the theory 
so that the singular points are at $u=1$ and $u=-1$.
Suppose that the singularity at $u=1$ is produced
by a state $(n_e ,n_m)$ in the $(\bf 2,\bf 1)$ of
$SU(2)\times SU(2)$ (this fixes our convention for the order
of the $SU(2)$ factors of $Spin(4)$). Thus, $n_e$ is even
and $n_m$ is odd, see Section 2.2.
The associated monodromy matrix is given by \div\ with $d=2$.
According to the discussion in Section 2.2, the state becoming
massless at $u=-1$ must then be $(n_e\pm n_m,n_m)$. Thus it lies in the
$(\bf 1,\bf 2)$ representation, in accordance with the fact
that \ZZ\ contains the
$\rho $ transformation.
The general consistency condition
for the monodromy group is 
$$M_{(n_e,n_m),2}\, M_{(n_e +n_m,n_m),2}=M_{\infty}=
M_{(n_e -n_m,n_m),2}\, M_{(n_e,n_m),2},
\eqn\qi $$
and this implies $n_m=\pm 1$. The electric charge is left arbitrary,
reflecting the possibility of ``democracy'' transformations,
and we choose $n_e=0$. 
This shows that, in this case again, the ``minimal'' choice of [\SWII ] is
actually the only possible one.

Now, one may note that the monodromies are exactly the same as in
the case of the $N_f=0$ theory in the old conventions 
of [\SW ] for  $a$ and the
electric charge.
This, together with the asymptotics \dii , completely determines the solution
to be
$$\eqalign{
\adeux(u)&= a(u)\cr
\add(u)&={1\over 2} \ad(u) \ , \cr }
\eqn\qii$$
where the functions $a(u)$ and $\ad(u)$ are defined by \dvi . It 
follows that $\add/\adeux = \ad/2a$ and that the curve of marginal
stability ${\cal C}_2$ for $N_f=2$ is the same as for $N_f=0$
(see Fig. 2). Furthermore, the
\ZZ\ symmetry is implemented formally exactly 
in the same way as in the case $N_f=0$ discussed in [\FB ].

The analysis of the spectrum can thus be done essentially without
modification. In the weak coupling spectrum
we now have the W bosons $(2,0)$ which
are singlets of $SU(2)\times SU(2)$ and the elementary quarks 
$(1,0)$ in the $(\bf 2,\bf 2)$ (which is nothing but the
defining representation of $SO(4)$), in addition to 
the dyons $(n,1)$. The latter are in the $(\bf 2,\bf 1)$ or in the
$(\bf 1,\bf 2)$ according to whether $n$ is even or odd.

In the strong-coupling region, one must introduce two different
descriptions
of the same section $p$ over $E_2$, one for
$\IM u>0$ (region ${\cal S}_{S+}$) and one for
$\IM u<0$ (region ${\cal S}_{S-}$).
We have
$$p\equiv (n_e,n_m)\in {\cal S}_{S+} \Longleftrightarrow 
p\equiv (n_e, n_m+2n_e) \in {\cal S}_{S-}\ .
\eqn\qii $$
Note that this transformation law is compatible with the
semi-classical formula for the constraints on the representations,
since $n_m+3n_e=n_m+n_e$ mod 2. This is related to the fact that
here the strong-coupling region is separated into only two pieces,
as already explained in Section 3.
One then shows that the states must come in \ZZ\ pairs which are
described, for example in ${\cal S}_{S+}$, as
$\big\{ \pm (n_e,n_m),\pm (n_e+n_m,-2n_e-n_m)\big\} $.
Note that the two $SU(2)$ factors of the flavour group are interchanged
for the two members of a given pair, as it should be.
Finally, using our by now standard argumentation, one can show that
there is only one \ZZ\ pair in the strong-coupling spectrum
containing
the monopole $\pm (0,1)$ and the dyon $\pm (\pm 1,1)$.

 $$\eqalign{
{\cal S}_{S+}&=\left\{\pm (1,-1)\ ,\ \pm (0,1)\right\}
=M_1{\cal S}_{S-} \cr
{\cal S}_{S-}&=\left\{\pm (1,1)\ ,\ \pm (0,1)\right\}
=M_1^{-1}{\cal S}_{S+}\ . \cr}
\eqn\qiii $$

\endpage

\section{Disintegrations}

We first list the predicted decay reactions, which here are
completely fixed looking only at the electric and magnetic
quantum numbers. We perform our analysis in ${\cal S}_{S+}$.
$$\eqalign{
\pm\pmatrix{1\cr 0\cr} &\,\,\longleftrightarrow\,\, \pm\pmatrix{1\cr -1\cr}
\pm\pmatrix{0\cr 1\cr}\cr
\pm\pmatrix{2\cr 0\cr} &\,\,\longleftrightarrow\,\, \pm 2\times\pmatrix{1\cr -1\cr}
\pm 2\times\pmatrix{0\cr 1\cr}\cr
\pm\pmatrix{n\cr 1\cr} &\,\,\longleftrightarrow\,\, \pm n\times\pmatrix{1\cr -1\cr}
\pm (n+1)\times\pmatrix{0\cr 1\cr}\cr.}
\eqn\qiv $$
One could now check that these reactions are compatible with mass conservation
and spin. This is very similar to what was already done in
[\FB ]. Here we focus on the flavour quantum numbers.
Recall that $\pm (1,-1)$ is in $(\bf 1,\bf 2)$ and that
$(0,1)$ is in $(\bf 2,\bf 1)$.
The decay of the quark is possible since the latter transforms in
$(\bf 2,\bf 2)$ and we have $(\bf 1,\bf 2)\otimes (\bf 2,\bf 1)
=(\bf 2,\bf 2)$. The decay of the W is also possible since
$(\bf 2,\bf 2)\otimes (\bf 2,\bf 2)$ contains the trivial 
representation.

What about the dyons? First note that ${\bf 2}^{\otimes n}$ contains
only $SU(2)$ representations of integer spin when $n$ is even,
and of half-integer spin when $n$ is odd. Thus 
$({\bf 2}^{\otimes (n+1)},{\bf 2}^{\otimes n})$ will contain
$(\bf 2,\bf 1)$ but not $(\bf 1,\bf 2)$ if $n$ is even, and 
it will contain $(\bf 1,\bf 2)$ but not $(\bf 2,\bf 1)$ if
$n$ is odd. But this is exactly what we need for the disintegration
of the dyons to be possible, since $\pm (n,1)$ is in
$(\bf 2,\bf 1)$ or in $(\bf 1,\bf 2)$ when $n$ is even
or odd.

\endpage

\chapter{Three flavours of quarks}

\section{The structure of the singularities and the global analytic structure}

{}From the analysis of [\SWII ], we know that we have two
singularities in the moduli space, one of them due to a massless
BPS state $(n_e,n_m)$ which is a singlet of the flavour group
$SU(4)$ (and thus $2n_e+n_m=0$ mod 4), 
and the other due to a state $(\tilde n_e,\tilde n_m)$ in
the defining representation $\bf 4$ of $SU(4)$ ($2\tilde n_e
+\tilde n_m=1$ mod 4). As there is no symmetry acting on the
Coulomb branch (except CP), the states $(n_e,n_m)$ and
$(\tilde n_e,\tilde n_m)$ are a priori unrelated. Moreover, at the
expense of 
shifting $u$, one can suppose that
$(\tilde n_e,\tilde n_m)$ is massless
at $u=0$ and $(n_e,n_m)$ is massless at $u=1/4$ (the scale is then
$\Lambda _3^2=64$).
The consistency condition for the monodromy group is
$$M_{(n_e,n_m),1}\, M_{(\tilde n_e +\tilde n_m,\tilde n_m),4}=M_{\infty}=
M_{(\tilde n_e ,\tilde n_m),4}\, M_{(n_e,n_m),1}.
\eqn\cons $$
This leads to an intricate system of algebraic equations, which 
have an invariance corresponding to the democracy transformations. 
We did not
try to solve this system explicitly. Instead, we give an argument
which does not require any computations and which fits the 
line of reasoning promoted in this work. 

In the problem at hand, we should again have a curve ${\cal C}_3$ of marginal
stability, diffeomorphic to a circle, passing through the singularities
(but we do not check the validity of this assertion for a general
structure of the singularities).
The analytic structure will be similar to the other cases where
only two singularities are present ($N_f=0,2$).
Because of the monodromy at infinity, $a_D/a$ will take all values
in the real interval $\lbrack\tilde n_e/\tilde n_m ,\tilde n_e/\tilde n_m
+1\rbrack =\lbrack r,r+1\rbrack $ along ${\cal C}_3$. Now, we know since the work
in [\SET ,\GAU ] that all the dyons $\pm (2n+1,2)$ with magnetic charge
two do belong to the semi-classical spectrum, and are singlets of
$SU(4)$. Since there is always an integer $n$ such that
$(2n+1)/2\in [r,r+1]$, we see that this state must become massless
somewhere on the curve. 
Of course this state is $(n_e,n_m)$. The choice of an odd integer
$n_e$ is then purely conventional, and we will take
$(n_e,n_m)=\pm (-1,2)$. Then one may return to \cons\ and show that
necessarily $(\tilde n_e ,\tilde n_m)=\pm (-1,1)$. This conclude
our analysis of the structure of the singularities,
and shows that the choice proposed in [\SWII ] is again the only
possible one.

The elliptic curve is in the case at hand [\SWII]
$$y^2=(x-u)(x^2-x+u)
\eqn\qi$$
and ${\rm d}\adt/{\rm d} u$ and ${\rm d}\at/ {\rm d} u$
are given by the corresponding period integrals. 
With appropriate contours $\gamma_i$ on the elliptic curve, $\adt$
and $\at$ are then given by
$${\sqrt{2}\over 8\pi} \oint_{\gamma_i} {\rm d}x 
{2u-x\over \sqrt{(x-u)(x^2-x+u)} } \ .
\eqn\qiii$$
It is straightforward to show [\IY] that $\adt$ and $\at$ given by the
integrals \qiii\ satisfy the differential equation
$$\left[ z(1-z) {{\rm d}^2\over {\rm d} z^2 } -{1\over 4}\right]
\pmatrix{ \adt\cr \at\cr} = 0\ ,
\eqn\qiv$$
where $z$ can be chosen to be $4u$ or equally $1-4u$.
This is the hypergeometric differential equation with $a=b=-{1\over 2}$,
and $c=0$. We observe that this is exactly the same equation as for
$N_f=0$, except that there $z={u+1\over 2}$. Hence, we immediately see
that  two independent solutions to \qiv\ are given by $\ad(8u-1)$ and
$a(8u-1)$ with $\ad$ and $a$ defined in \dvi. 

It remains to determine the correct normalisations and linear combinations.
{}From the required asymptotics \dii\ as $u\to \infty$ together with \dviia\ 
one concludes
$\at(u)={\sqrt{2}\over 4} a(8u-1)$ and
$$\adt(u)={\sqrt{2}\over 16} \Bigl[ \ad(8u-1)-\gamma a(8u-1) \Bigr] \ .
\eqn\qvi$$
To determine the constant $\gamma$, we
simply require to have a massless 
dyon $(-1,2)$ at $u=1/4$, that is $a(1/4)+2a_D(1/4)=0$. This leads to
$\gamma =2$ and then
$$\eqalign{
\adt(u) &= {\sqrt{2}\over 16}  \Bigl[ \ad(8u-1)- 2\, a(8u-1) \Bigr] \cr
        &= {\sqrt{2}\over 4} \biggl[  i \Bigl(u-{1\over 4}\Bigr) 
 F\Bigl({1\over 2}\raise 2pt\hbox{,}{1\over 2}\raise 2pt\hbox{,}2;1-4u\Bigr) 
-\sqrt{u}  F\Bigl(-{1\over 2}\raise 2pt\hbox{,}{1\over 2}\raise 2pt\hbox{,}1;{1\over 4 u}\Bigr) \biggr] \cr
\at(u)  &=  {\sqrt{2}\over 4} a\bigl(8u-1\bigr) = \sqrt{{u\over 2}}\, 
 F\Bigl(-{1\over 2}\raise 2pt\hbox{,}{1\over 2}\raise 2pt\hbox{,}1;{1\over 4 u}\Bigr) \ . \cr
}
\eqn\dvii$$
The positions of the branch cuts are shown in Fig. 9.
\vskip 3.mm
\fig{The branch cuts of $\adt(u)$ and $\at(u)$}{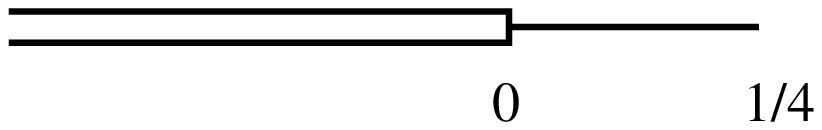}{6cm}
\figlabel\figiv
\vskip 2.mm
The various asymptotics of $\adt$ and $\at$ at the singularities are 
$$\left.\eqalign{
\adt(u)&\simeq {i\over 4\pi} \sqrt{2u}\left[ \log u +6 \log 2-2 + i\pi \right] \cr
\at(u)&\simeq {1\over 2} \sqrt{2u} } \quad \right\} \quad {\rm as\ } u\to\infty$$
$$\left.\eqalign{
\adt(u)&\simeq  {\sqrt{2}\over 16 \pi}  \left[ -4 +(4u-1)\log(4u-1) 
+(4u-1)(1-4 \log2 + i\pi) \right] \cr
\at(u)&\simeq   {\sqrt{2}\over 8 \pi}  \left[ 4 -(4u-1)\log(4u-1) 
-(4u-1)(1-4 \log2) \right]   } \quad  \right\} \quad {\rm as\ } u\to {1\over 4}$$
$$\left.\eqalign{
\adt(u)&\simeq   {\sqrt{2}\over 16 \pi} i (1+\e) \left[ -4 -4u\log 4u + 4u (1+4\log
2) \right] -{\sqrt{2}\over 4} u   \cr
\at(u)&\simeq   {\sqrt{2}\over 8 \pi} i \e \left[ 4 +4u\log 4u - 4u (1+4\log 2) 
\right] + {\sqrt{2}\over 2} u  } \quad  \right\} \quad {\rm as\ } u\to 0 \ ,
\eqn\qviii$$
where again $\epsilon$ is the sign of $\IM\, u$. It is then straightforward to
determine the monodromy matrices for analytic continuation of 
$(\adt,\at )$ around these three singular points:
$$ M_\infty=\pmatrix{-1 &\hfill 1\cr\hfill 0 &-1\cr}\ , \  
M_{1/4}=\pmatrix{\hfill 3 &\hfill 1\cr -4 &-1\cr}\ , \  
M_0=\pmatrix{\hfill 1&0\cr -4&1\cr}\ , \  
M_{0}'=\pmatrix{\hfill 5 &\hfill 4\cr -4 &-3\cr}\ , \  
\eqn\dix$$
where $M_0$ is to be used if the monodromy around $u=0$ is computed with a basepoint
in the lower half $u$-plane ($\e=-1$) and $M_{0}'$ if the basepoint is in the upper
half $u$-plane ($\e=+1$). 

\section{The curve  of marginal stability}

Let's now discuss the curve \C\ of marginal stability. Again this is the set of all $u$
such that $\wt\equiv \adt / \at$ is real. Since
$$\wt(u)\equiv {\adt\over \at}={1\over 4} \left[ {\ad(8u-1)\over a(8u-1)} -2 \right]
\equiv {1\over 4} \left[ w(8u-1)-2 \right] \ ,
\eqn\qxi$$
this curve is a rescaled and shifted copy of the curve for $N_f=0$, which is almost
an ellipse, see Fig. 10.
\vskip 2.mm
\fig{The curve of marginal stability for $N_f=3$}{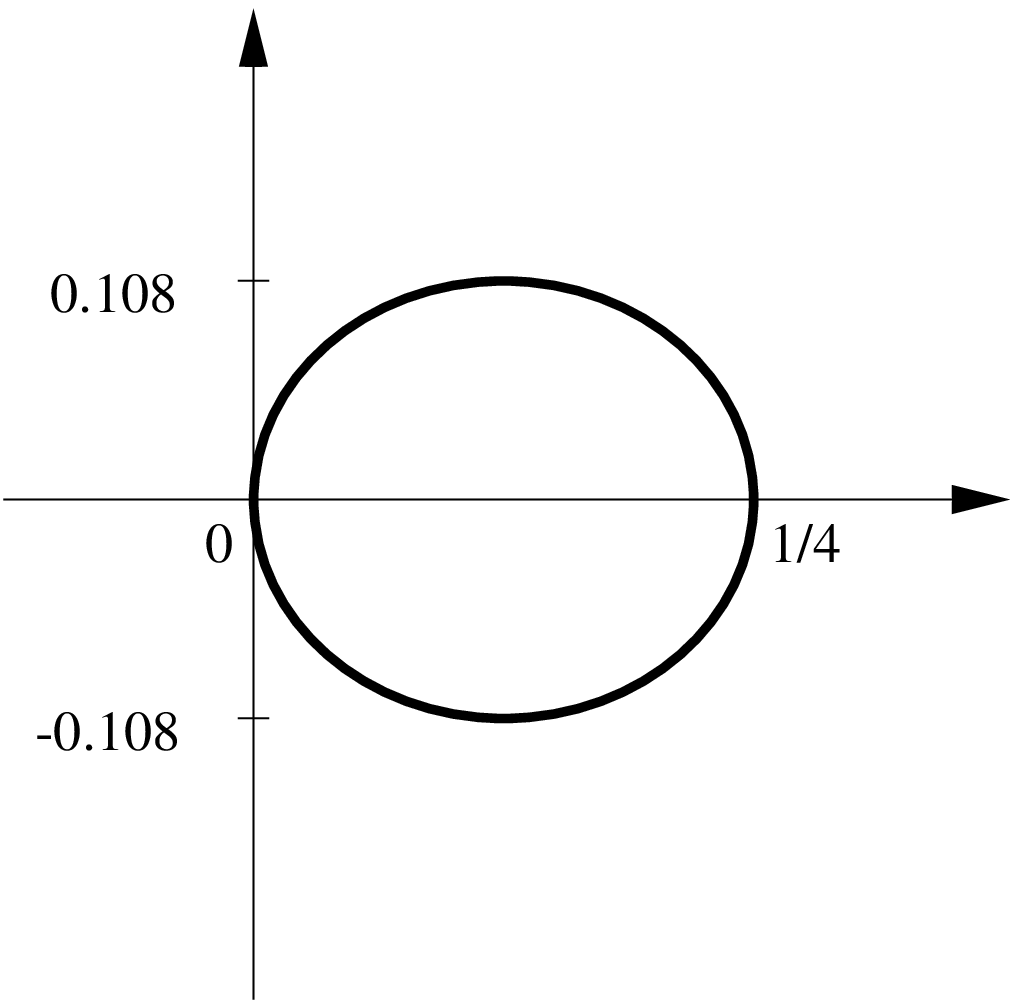}{6cm}
\figlabel\figv
\vskip 2.mm
If we again call \CP\ and \CM\ the parts of the curve in the upper half and lower
half $u$-plane, then
$$\wt \in [-1,-{1\over 2}] \ {\rm on} \ {\cal C^+} \ , \quad 
\wt \in [-{1\over 2},0] \ {\rm on} \ {\cal C^-} 
\eqn\qxii$$
with $\wt(0+i\e)=-1,\ \wt({1\over 4})=-{1\over 2}$ and $\wt(0-i\e)=0$, as follows
immediately from the corresponding properties [\FB] of $w$ we recalled in Sect. 2.1, 
but also from \qviii.

\section{The weak and strong-coupling spectra}

Next, we determine the weak and strong-coupling spectra \sw\ and \ss, corresponding to
the regions \rw\ and \rs\ outside and inside the curve \C.
In all of \rw, a BPS state can be described by a unique couple of
integers \pp. However, similarly to what happened for $N_f=1$ or $2$, 
in the strong-coupling region \rs\  one
needs to introduce two descriptions \pp\ and $(\tilde n_e, \tilde n_m)$ for $\IM u>0$
(in \rsp) and $\IM u<0$ (in \rsm), for the {\it same} BPS state $p$. 
%
%
%
%
%
By exactly the same argument as for the other cases, this time one finds
$$p\equiv (n_e,n_m)\in {\cal S}_{S+}
\quad \Longleftrightarrow \quad p\equiv (-n_e-n_m,4n_e+3n_m) \in {\cal S}_{S-}
\ .\eqn\qxv$$
Now we are in a position to determine the strong-coupling spectrum. While the
semi-classical monodromy $M_\infty$ must be a symmetry of the weak-coupling spectrum,
this need not and will not be the case for the strong-coupling spectrum.
Also, here we do not have any broken global quantum symmetry \Z $_k$ at our disposal.
But this is actually not needed. As is by now familiar,
a BPS state described by \pp$\in$\ssp\
will become massless somewhere on \CP\ if 
$\displaystyle{{n_e\over n_m}}\equiv r \in [-1,-{1\over 2}]$. 
This same BPS state is described  by $(\tilde n_e, \tilde n_m)\in$\ssm, and it
will be massless somewhere on \CM\ if 
$${\tilde n_e\over \tilde n_m}= - {n_e+n_m\over 4n_e+3n_m}
= - {r+1\over 4r+3}\equiv \varphi(r)
\eqn\qxvi$$
is such that $ \varphi(r)\in [-{1\over 2},0]$. But it is easy to see that the
function $ \varphi(r)$ is such that either $r\in [-1,-{1\over 2}]$ (and then 
$ \varphi(r)\notin (-{1\over 2},0)$) or 
 $ \varphi(r)\in [-{1\over 2},0]$ (for $r\notin (-1,-{1\over 2})$). Hence any BPS
state in the strong-coupling spectrum becomes massless somewhere on the curve \C.
But a massless BPS state always leads to a singularity. Since there
are only two singularities, at $u=0$ and $u={1\over 4}$, the only states in the
strong-coupling spectrum are the two states (together with their anti-particle)
responsible for the singularities, namely the magnetic monopole, described as
$(0,1)\in$\ssm\ or $(1,-1)\in {\cal S}_{S+}$, and the dyon $(-1,2)\in {\cal S}_{S\pm}$:
$$\eqalign{
{\cal S}_{S+}&=\left\{ \pm (-1,2)\ , \ \pm (1,-1) \right\} \cr
{\cal S}_{S-}&=\left\{ \pm (-1,2)\ , \ \pm (0,1)  \right\} \ . \cr }
\eqn\qxvii$$

It remains to determine the weak-coupling spectrum \sw. As already mentioned, the
only symmetry here is the monodromy around $u=\infty$, and one must have
$$M_\infty {\cal S}_W={\cal S}_W \ .
\eqn\qxviii$$
Since the states that become massless at $u=0$ and $u={1\over 4}$ must be present in
the weak and strong-coupling spectrum, we know that $(-1,2), (0,1) \in$\sw. Applying
$M_\infty^k\, ,\ k\in \Z$, to these states one obtains all dyons $(n,1)$ and
$(2n+1,2), \ n\in \Z$. 
The dyons $\pm (2n+1,2)$ obtained in this way are all singlets
of $SU(4)$ since of course $M_{\infty}$ does not change the flavour
representations. One might ask whether it is possible  to have
dyons $\pm (2n+1,2)'$ which are not singlets of $SU(4)$ (note that
for $|n_m|=1$ we are automatically in a spinor representation and thus
a similar question would be irrelevant). But if this were the case,
$\pm (-1,2)'$ would be massless at $u=1/4$ together with
$\pm (-1,2)$, and thus $M_{1/4}$ would be changed, which is excluded.
Of course, the elementary states, namely the quarks $(1,0)$
and W-bosons $(2,0)$ 
are also present in \sw. Can one have dyons with magnetic charge
$|n_m| \ge 3$ ? The answer is no. 
Suppose there were such a dyon \pp$\in$\sw\ with
$n_m\ge 3$. Then \sw\ would contain all dyons $(-)^k M_\infty^k
(n_e,n_m)=(n_e-kn_m,n_m)$. There is always a $k\in \Z$ such that 
$\displaystyle{{n_e-kn_m\over n_m}={n_e\over n_m}-k\in [-1,0]}$. 
One would then conclude that $(n_e-kn_m,n_m)$
becomes massless somewhere on \C. But there 
are no massless states with $n_m\ge 3$.
Hence we conclude that \sw\ cannot contain states with $n_m\ge 3$, and that
$${\cal S}_W
=\left\{ \pm (1,0), \pm (2,0), \pm (n,1), \pm (2n+1,2), \ n\in \Z \right\} \ .
\eqn\qxix$$

\section{Disintegrations}

As in the case $N_f=2$, the predicted decay reactions are
completely fixed when one takes into account the conservation of 
the electric and magnetic charge. We perform our analysis
in ${\cal S}_{S-}$.
$$\eqalign{
\pm\pmatrix{1\cr 0\cr} &\,\,\longleftrightarrow \,\,\pm 2\times \pmatrix{0\cr 1\cr}
\pm\pmatrix{1\cr -2\cr}\cr
\pm\pmatrix{2\cr 0\cr} &\,\,\longleftrightarrow \,\,\pm 4\times\pmatrix{0\cr 1\cr}
\pm 2\times\pmatrix{1\cr -2\cr}\cr
\pm\pmatrix{n\cr 1\cr} &\,\,\longleftrightarrow \,\,\pm (2n+1)\times
\pmatrix{0\cr 1\cr}
\pm n\times\pmatrix{1\cr -2\cr}\cr
\pm\pmatrix{2n+1\cr 2\cr} &\,\,\longleftrightarrow \,\,\pm 4(n+1)\times
\pmatrix{0\cr 1\cr}
\pm (2n+1)\times\pmatrix{1\cr -2\cr}\cr}.
\eqn\qx $$
We focus on the flavour quantum numbers, all other consistency conditions being
satisfied, similarly
to the previous cases. 
First we present some
Clebsch-Gordan series for tensor products of the $\bf 4$ of
$SU(4)$ which actually will be enough to obtain the desired results.
These are
$$\eqalign{
{\bf 4}^{\otimes 2} &={\bf 6}\oplus {\bf 10}\cr
{\bf 4}^{\otimes 3} &={\bf 20}_a\oplus 2\times {\bf 20}_b
\oplus \overline{{\bf 4}}\cr
{\bf 4}^{\otimes 4} &={\bf 1}\oplus 3\times {\bf 15}\oplus 2\times
{\bf 20}_c \oplus {\bf 35}\oplus 3\times {\bf 45}\cr }.
\eqn\new $$
We recall that the monopole $(0,1)$ is in the ${\bf 4}$ and its anti-particle
$(0,-1)$ is in the complex conjugate representation 
$\overline{\bf 4}$,
while $\pm (-1,2)$ is a singlet.
Now it is clear that the decay reactions for the elementary quarks
$(+1,0)$ are possible since ${\bf 4}\otimes {\bf 4}$ contains ${\bf 6}$ which
is the defining representation of $SO(6)$ of which the quarks
form a multiplet. Simply taking the complex conjugate of the
Clebsch-Gordan series show that the corresponding reaction for the
anti-particles is also possible (this is always the case, and we will
thus only study the decays of $(+2,0)$, $(n,+1)$ and $(2n+1,+2)$ in
the following). 
The reaction is possible for the W$^+$ because ${\bf 1}$ is contained
in ${\bf 4}^{\otimes 4}$.
Moreover, for $n$ even, ${\bf 4}^{\otimes (2n+1)}=({\bf 4}^{\otimes 4})
^{\otimes (n/2)}\otimes {\bf 4}$ contains $\bf 4$ as 
${\bf 4}^{\otimes 4}$ contains $\bf 1$, and thus the decay
of the dyon of unit magnetic charge and even electric charge is
consistent. The same is true when the electric charge is odd,
because then ${\bf 4}^{\otimes (2n+1)}=({\bf 4}^{\otimes 4})
^{\otimes (n-1)/2}\otimes {\bf 4}^{\otimes 3}$ and
${\bf 4}^{\otimes 3}$ contains $\overline{\bf 4}$.
We end the discussion with the decay of the dyons of magnetic
charge 2. Here, the desired result follows from the fact that
${\bf 4}^{\otimes 4(n+1)}$ contains the trivial representation in its
Clebsch-Gordan series.

\chapter{Conclusions}

In this paper, we have shown that the results of [\FB] for $N_f=0$ can be extended to
the general asymptotically free $N=2$
supersymmetric Yang-Mills theories with gauge group
$SU(2)$ when massless hypermultiplets
are present. In all cases, we obtained the global structure of the $SL(2,\Z)$
bundles,
showed the existence and studied the 
properties of the curves of marginal stability,
separating the strong from the so-called 
weak-coupling regions. For $N_f=1$ and $N_f=2$, we used the discrete
global $\Z _3$ or $\Z _2$ symmetry on the Coulomb
branch of the moduli space and showed that the states in the strong-coupling
region must come 
in multiplets of the corresponding 
spontaneously broken discrete global symmetry
$\Z_{4(4-N_f)}$. 
We saw for $N_f=3$ that the existence of such a global symmetry
was not crucial. Even for $N_f=1$ and 2 one might not have used it
explicitly: it is really the global structure of the moduli space 
which gives all the constraints we need. Of course this global structure
is strongly constrained itself by the global symmetry when it exists.

In all cases we determined 
the strong and weak-coupling spectra of BPS
states completely and rigorously. 
The strong-coupling spectra contain precisely only
those states that are responsible for the 
singularities at finite points of the Coulomb
branch. We showed that the discontinuities of 
the spectra across the curves of marginal
stability lead to disintegrations that 
always are perfectly consistent with the
conservation of mass, electric, magnetic and flavour charges. 
As a byproduct, we also showed
that the electric and magnetic quantum numbers of the massless states 
at the singularities proposed by Seiberg 
and Witten are the only possible ones.

\ack
One of us (A.B.) is grateful to D. L\"ust and the members of the Theoretical Physics
group at the Humboldt University in Berlin for their welcoming hospitality while
part of this work was carried out there.
\refout
\end